\documentclass{article}
\usepackage[utf8]{inputenc}
\usepackage{authblk}
\usepackage{indentfirst}
\usepackage{hyperref}
\usepackage{longtable}
\usepackage{setspace}
\usepackage[margin=1in]{geometry}
\usepackage{graphicx}
\graphicspath{ {./figures/} }
\usepackage{subcaption}
\usepackage{amsmath}
\usepackage{lineno}
\usepackage{multirow}
\usepackage{color,soul}
\usepackage[version=3]{mhchem}

\usepackage[biblabel]{cite}

\title{Identifying intercalation electrodes for electrochemical ion recovery and desalination}

\author[1]{Sougat Purohit}
\author[2,3]{Andreas Kuhlmann}
\author[2,3]{Marten Huck}
\author[2,3]{Hans-Georg Steinr\"{u}ck}
\author[1,*]{Gopalakrishnan Sai Gautam}

\affil[1]{Department of Materials Engineering, Indian Institute of Science, Bengaluru, 560012, India}
\affil[2]{Institute for a sustainable Hydrogen Economy (IHE-1), Forschungszentrum J\"{u}lich GmbH, An der Deutschen Welle 7a, J\"{u}lich 52428, Germany}
\affil[3]{Institute of Physical Chemistry, Rheinisch-Westf\"{a}lische Technische Hochschule Aachen, Landoltweg 2, Aachen 52074, Germany}
\affil[*]{Email: \href{mailto:saigautamg@iisc.ac.in}{saigautamg@iisc.ac.in}}

\date{}

\begin{document}

\maketitle

\begin{abstract}
\noindent Water sources serve as a requirement for human consumption, industrial, and agricultural needs and a resource for valuable ions. Appropriate treatment of water that captures ions from the aqueous solution can give rise to suitable `products', such as critical ions (such as Li$^+$) in selective ion recovery and usable water in the case of desalination (i.e., removal of non-essential ions). Among myriad ion capture techniques, we focus on electrochemical methods using intercalation electrodes, an energy-efficient approach capable of both selectively extracting specific ions or non-selectively extracting multiple ions from diverse water sources. However, practical realization of electrochemical ion capture requires the use of robust cation intercalation electrodes, with the state-of-the-art olivine-FePO$_4$ and spinel-Mn$_2$O$_4$ exhibiting several shortcomings. Thus, we develop a high-throughput computational workflow to identify potential intercalation electrodes for both ion recovery and desalination from a pool of 20,414 Li, Mg, Ca, Na, and/or K-containing structures from the Materials Project database. Based on the construction of Pourbaix diagrams relevant for electrodes operating under seawater conditions, modifying criteria used for defining metastability in seawater, computing average intercalation voltages and the extent of metastability of electrode structures, and computational graphs to encode ionic selectivity, we identify a total of 11 promising electrodes for selective ion recovery and one possible desalination electrode. The promising materials identified in this work include Fe$_2$(MoO$_4$)$_3$, Li$_3$Fe$_2$(PO$_4$)$_3$, LiFeP$_2$O$_7$ for selective Li-recovery. Our study provides thermodynamic insights, qualitative trends, and a promising list of intercalation electrodes for experimental testing that should advance the field of electrochemical ion recovery.
\end{abstract}


\section{Introduction}
\noindent Conserving potable water and managing industrial effluents are crucial in an era of increasing salinity of freshwater resources caused by rising sea levels and anthropogenic activities \cite{kaushal2023anthropogenic}. Desalination, the process of separating ions from water, can turn the challenge of increasing salinity into economic opportunities by sustainably extracting desired ions normally found in saline water (e.g., Li, Mg, K, etc.) and/or by producing clean water \cite{khanzada2026desalination, o2024analysis}. Note that the objective of desalination is generally to produce cleaner water suitable for drinking, irrigation, or disposal post industrial usage. On the other hand, selective ion recovery \cite{gamaethiralalage2021recent, srimuk2020charge} involves directly extracting a specific ion from salt water and/or waste water sources. Importantly, both desalination and selective ion recovery rely on similar thermodynamics as their working principles, with differences in the eventual `product'.

With respect to ion recovery, one critical element that can be sustainably extracted from water-based sources is Li, the electroactive component of the most widely used energy storage technology, namely the Li-ion battery (LIB). The global push for renewable energy (and its associated storage) is expected to disrupt Li supply in the near future due to continuously increasing demand and uneven distribution of Li worldwide \cite{meng2025technological, olivetti2017lithium, battistel2020electrochemical}. For instance, the Li-rich brines of South America constitute ~42\% of the global Li reserves \cite{USGS-Li2026} and rely on slow and water-intensive solar thermal evaporation followed by chemical separation of Li from the leftover salt for Li extraction \cite{vera2023environmental}. In this context, direct Li extraction (DLE~\cite{pasta2012batteries,xu2025membrane, kong2025electro}) from aqueous sources provides a sustainable method to extract Li from non-conventional sources like geothermal brines \cite{stringfellow2021technology}, seawater \cite{bardi2010extracting, liu2020lithium, li2021continuous}, wastewater from battery recycling \cite{kim2018electrochemical, fan2020sustainable}, as well as from effluents of oil and gas plants~\cite{kumar2019lithium}. While DLE from these sources involves selectively capturing Li$^+$ in the presence of competing cations, such as Na$^+$, K$^+$, Mg$^{2+}$, and Ca$^{2+}$ \cite{li2021continuous, li2024lithium, yan2022defining, kong2025electro, peng2024extreme}, desalination requires capturing all (possible) ions, including Li$^+$. Note that the extent to which ``all" ions must be removed for desalination also depends on the overall composition of the water source, the specific use case (drinking, industrial or agricultural) and the respective regulatory specifications \cite{EU2020Directive2184, Saha2026IrrigationWater}.

Traditionally, desalination involves energy-intensive thermal evaporation of water, leaving behind salts \cite{wang2021electrochemical, zheng2021large}, with the recovery of clean water via condensation. However, the most widely used industrial technique for seawater desalination is reverse osmosis (RO) \cite{elimelech2011future, dudchenko2021high, liu2024pressure, alnajdi2024practical, jones2019state}, where pressure gradients force water through a selectively permeable membrane that separates water from dissolved ions. One limitation of RO processes is the expensive semi-permeable membrane, which can be susceptible to damage and fouling upon repeated use~{\cite{niemann2023x, bone2020advanced}}. Moreover, RO by construction is not selective to a given element, such as Li, making it unsuitable for direct mineral extraction from aqueous sources. Thus, it is important to develop common technological platforms (or frameworks) that can be leveraged for either ion recovery or desalination, depending on the water source and the desired product.

Other techniques of ion removal from aqueous sources include hydrometallurgical methods involving the addition of precipitating solvents \cite{bajpayee2011very, guo2021ionic, dach2026novel}, chromatography \cite{ding2020two, dong2021bridging, mattocks2023enhanced}, and electrochemical methods \cite{pasta2012batteries, pasta2012desalination, knust2013electrochemically, jeon2013desalination, suss2015water, srimuk2020charge, do2023embedded, xu2024pseudo}. Out of these techniques, we focus on  electrochemical methods since they have the potential to act as common technological framework for both ion recovery and desalination, are typically less energy-intensive, and exhibit the versatility to handle various solutions (e.g., different concentrations of salts in solution). Notably, the choice of electrodes and their selectivity to various ions, applied potentials, and other operating conditions (temperatures, flow rates of solutions, etc.) can be tuned such that a given electrochemical setup can be used for either ion recovery (for mineral extraction) or desalination (for clean water generation), as schematically displayed in \textbf{Figure~S1} of the supporting information (SI). Through the rest of the manuscript, we will employ the terminology, `ion capture' to refer to either desalination or ion recovery.

Among the multiple variants of electrochemical techniques, we focus on employing intercalation electrodes that can extract ion(s) from a solution, a process with operating principles similar to LIBs \cite{battistel2020electrochemical, xiong2022electrochemical, suss2018water, dunn2011electrical}. Intercalation electrodes work by absorbing ions (from the solution) into the available lattice (interstitial) sites in their structure under an applied potential/current, with cations usually intercalating better compared to the larger-sized and oppositely-charged anions (such as Cl$^-$). Here, we screen for materials that can be suitable for capturing cation(s), i.e., those that can act as negative electrodes in an electrochemical setup (\textbf{Figure~S1}), while we assume anions in the solution to be simultaneously captured by a capacitive or a conversion-type positive electrode \cite{xiong2022electrochemical, srimuk2020charge, battistel2020electrochemical}. The efficient capture of cation(s) depends on the properties of the intercalation electrodes, which makes ion capture performance a material-dependent property. Hence, to improve ion capture, promising intercalation electrodes and an improved understanding of the associated mechanisms and processes are required.

So far, intercalation electrodes that have been employed for ion capture from aqueous solutions include olivine-FePO$_4$ \cite{pasta2012batteries, trocoli2015nickel, liu2020lithium} and Mn$_2$O$_4$ (or MnO$_2$ equivalently) polymorphs, such as the spinel \cite{trocoli2017optimized, kim2018electrochemical, zhao2019highly, huck2026toward}. FePO$_4$ and Mn$_2$O$_4$ have been explored due to their ability to intercalate a variety of ions simultaneously and reversibly (for desalination) and/or their relatively high selectivity for Li$^+$ over other cations (for ion recovery). However, both FePO$_4$ and Mn$_2$O$_4$ electrodes are limited by different factors, such as the poor electronic conductivity of olivine-FePO$_4$ resulting in slow intercalation kinetics \cite{ling2021phosphate} and the poor structural stability of spinel-Mn$_2$O$_4$ due to Mn dissolution during cycling \cite{ouyang2009jahn, jiang2022facet}. Attempts to mitigate the such shortcomings include carbon coating the FePO$_4$ to improve the intercalation kinetics \cite{wang2021electrochemical} and doping Ni in Mn$_2$O$_4$ to improve its structural stability \cite{shang2021lini0}. Also, there have been attempts to make LiFePO$_4$ more selective towards Li$^+$  by using a TiO$_2$ coating \cite{liu2020lithium}. Overall, the literature has focused on employing morphological/structural/compositional modifications of olivine-FePO$_4$ and spinel-Mn$_2$O$_4$ for ion capture. As a complementary pathway to improve electrochemical ion capture, we propose a high-throughput computation-driven approach to explore different chemical and structural spaces, and identify novel and sustainable electrodes with intrinsically better properties. 

The cation intercalation electrodes in ion capture processes absorb/release Li$^+$ (or by extension, other cations), via the (de)intercalation mechanism with the application/removal of an external potential/current, which is identical to the operation of a positive electrode (cathode) in LIBs. Thus, the extensive research on LIB and beyond-LIB cathodes provides immense data to identify novel electrodes that can achieve better ion capture efficiencies than either olivine-FePO$_4$ or spinel-Mn$_2$O$_4$ based electrodes. Thus, in this work, we use the Materials Project (MP) database \cite{jain2013commentary}, built from high-throughput density functional theory (DFT \cite{hohenberg1964inhomogeneous, kohn1965self}) calculations as the primary data source, and apply material-specific screening criteria (as described below) to find promising electrode candidates for ion capture.

For a material to be useful as an ion capture electrode, there are constraints the material must satisfy, which are different from the constraints typically used for battery cathode selection \cite{ceder1998identification, hautier2011phosphates, deb2022critical}. For instance, LIBs and analogous battery technologies contain non-aqueous electrolytes with wider voltage stability windows ($\sim$3-4~V \cite{xu2014electrolytes, hall2018exploring, peljo2018electrochemical, yamada2019advances}) than water ($\sim$1.23~V \cite{suo2015water}) that allow for higher operating voltages. Thus, water being the medium for ion capture limits the range of applied potentials, versus standard hydrogen electrode (SHE), within which an electrode has to operate. Moreover, the potential dissolution of non-electroactive species (such as Mn$^{2+}$ \cite{hua2019structural, liu2026dual}) can detrimentally affect the long-term stability and cycle life of the electrode, making stability of the material in aqueous solutions an important criterion. However, information on the aqueous stability of materials is not readily available in any computational/experimental database. Additionally, we require materials with good ionic selectivity, i.e., intercalate only the desired ion among other ions present in the solution, for ion recovery, while we require electrodes with low selectivity for desalination. Thus, aqueous stability, `optimum' intercalation voltages, and ionic (non-)selectivity are important screening criteria for identifying novel electrodes for electrochemical ion capture. 

To determine the aqueous stability of potential electrode materials, we make use of DFT-based Pourbaix diagrams \cite{pourbaix1966atlas}, which are phase diagrams built with semi-grand canonical boundary conditions and plotted on an electrode potential (referred to as applied potential in this work) vs.~pH scale at a given temperature (usually 298~K) \cite{mccafferty2010thermodynamics}. Pourbaix diagrams have been previously employed in the high-throughput screening of photocatalysts for CO$_2$ reduction \cite{singh2019robust}, and in the identification of water-stable metals, intermetallics, and metal oxides \cite{zhang2025aqueous}. Moreover, recent works by Du et al.\mbox{\cite{du2025accelerating}} and Ruiz et al.\mbox{\cite{garay2026enriching}} extend the utility of Pourbaix diagrams for exploring interfacial properties and phase prevalence of materials. Similarly, ion capture presents an additional challenge of the electrode having to be stable in an aqueous solution containing additional non-electroactive elements (as compared to pure water), which can affect the aqueous stability of the electrode. Thus, we introduce a modified metastability formalism within our calculated Pourbaix diagrams that incorporates the presence of common ionic species in seawater-based brine sources, such as Cl, S, Na, and Mg, to evaluate the stability of potential electrodes in multi-component aqueous solutions. Note that the most critical cation to be captured in a desalination process is usually Na$^{+}$, given its large abundance in seawater. 

In this work, we utilize DFT-calculated Pourbaix diagrams, develop associated screening criteria, and construct computational graphs to identify promising intercalation electrodes for electrochemical ion capture. We introduce important modifications on the Pourbaix diagram construction to include ($i$) additional ions in the aqueous phase that affect the stability of electrodes; ($ii$) a re-defined electrode metastability criterion in seawater that is extendable to any desired aqueous composition; and ($iii$) intercalation voltages under operating conditions to identify promising electrodes. Overall, we screen 20,414 structures from MP and identify 11 promising electrodes for ion recovery (with high selectivity) and one desalination electrode (with low selectivity) that can operate in seawater. We provide detailed descriptions and schematics of all the candidates identified in this work, with a special emphasis on the candidates that share similarities with experimentally known structures. 

Notably, we identify pyrophosphate-LiFeP$_2$O$_7$, anti-NASICON-Fe$_2$(PO$_4$)$_3$, and garnet-like-Fe(MoO$_4$)$_2$ as the most promising Li-selective electrodes for Li-ion recovery. Additionally, we find 8 structures, including orthophosphates, alluaudites, spinels, perovskites, and hollandite-like structures that are selective towards one of the seawater-abundant cations, namely, Na, K, Mg, and Ca. Besides electrodes for ion recovery, we find 8 electrodes, that can intercalate two, three, or four of Li, Na, K, Mg, and Ca. For instance, we find novel (theoretical) electrodes that can simultaneously intercalate Ca$^{2+}$ and Mg$^{2+}$, namely, monoclinic-FePO$_4$ and triclinic-Fe$_2$(MoO$_4$)$_3$, with possible utility in water-softening. While we do not find any electrode that intercalates all five cations, olivine-FePO$_4$ is the most promising desalination electrode that we identify given that it can intercalate four (Li, Na, K, and Mg) out of five cations considered. Finally, our work provides a robust, systematic, computation-driven screening strategy and identifies intercalation electrodes for electrochemical ion capture, which should improve the associated process efficiencies in the near term and enable a successful transition to a sustainable water utilization and energy future.

\section{Methods}
\subsection{Workflow}
\label{sec:workflow}
\begin{figure}[h!]
\centering
\includegraphics[width=0.8\textwidth]{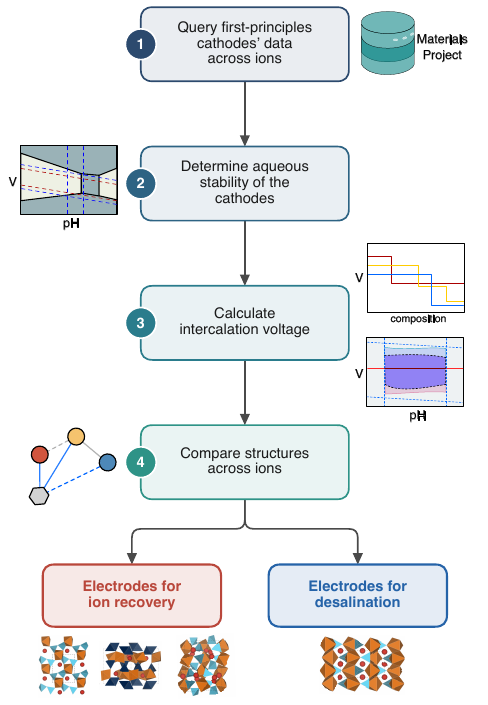}
\caption{Schematic of the workflow used in the current work, including obtaining DFT data, generating seawater Pourbaix diagrams and evaluating thermodynamic (meta)stability, calculating operating voltages, and identifying unique structural candidates that are suitable for electrochemical ion recovery and desalination.}
\label{fig:1}
\end{figure}

\noindent \textbf{Figure~{\ref{fig:1}}} provides a schematic of the workflow used in this work, involving four distinct steps. We start by querying first principles calculated data from MP for materials that can potentially intercalate Li, Na, K, Mg, and Ca (see \textbf{Section~{\ref{sec:query}}}). In the second step, we generate Pourbaix diagrams for all the queried electrode compositions by including the four most abundant elements present in seawater (besides H and O), namely, Cl, Na, S, and Mg, to quantify the aqueous (seawater) stability of the electrodes. We refer to the generated Pourbaix diagrams as the `seawater Pourbaix diagrams' (\textbf{Section~{\ref{sec:seawater-Pourbaix}}}). In the third step, for all electrode candidates that are (meta)stable in seawater (\textbf{Section~{\ref{sec:metastability}}}), we evaluate the thermodynamic average intercalation voltages and the extent to which the metastable phases can exist across the pH and potentials in seawater (\textbf{Section~{\ref{sec:voltages}}}). Finally, we compare the electrode-frameworks (\textbf{Section~{\ref{sec:comparison}}}) across the five different electroactive cations (i.e., Li, Na, K, Mg, and Ca), to identify frameworks that are selective to a single ion (for ion recovery) and structures that can intercalate multiple ions (for desalination). We make this step-wise workflow available as a \texttt{python} package named `seawater Pourbaix intercalation cation electrode screening' (\texttt{SPICES}) on the \href{https://github.com/sai-mat-group/SPICES}{GitHub} repository associated with this work.

\subsection{Data and querying process}
\label{sec:query}
\begin{figure}[h!]
\centering
\includegraphics[width=\textwidth]{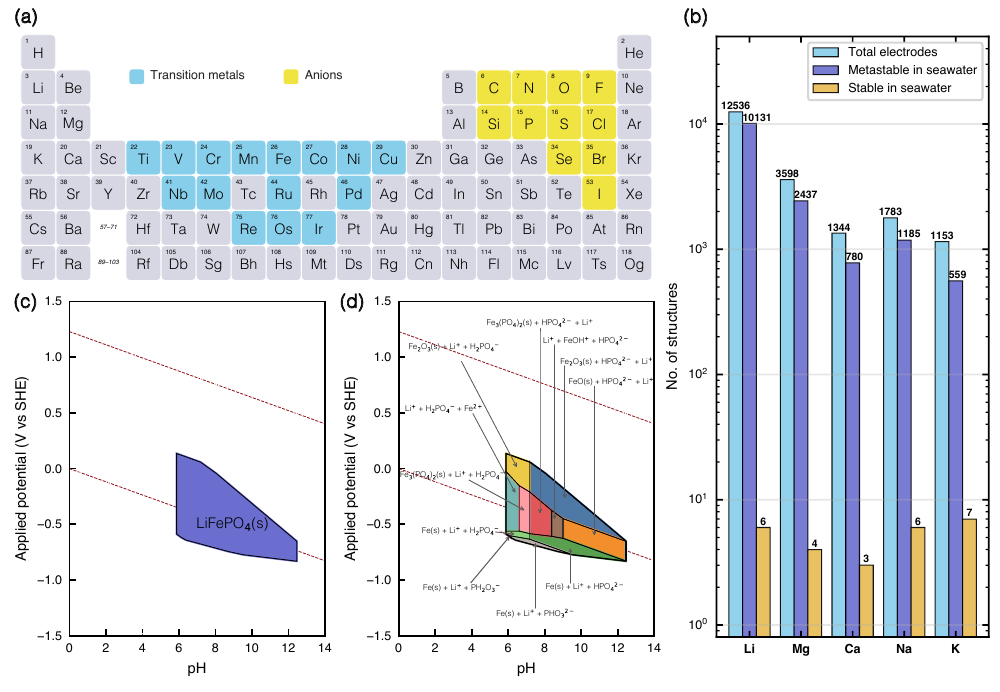}
\caption{(a) Periodic table snippet showing redox-active TMs (blue) and anionic (yellow) elements considered in our search for possible ion capture electrodes. (b) Bars representing the total number of queried (cyan), seawater metastable (blue), and seawater stable (yellow) electrode structures containing different electroactive ions. (c), (d) Pourbaix diagrams showing the thermodynamic stability of LiFePO$_4$ at aqueous concentrations of (c) [Li] = 0.75~M, [Fe] = 0.25~M and [P] = 0.33~M and (d) in the dilute limit of [Li] = [Fe] = [P] = $10^{-6}$~M. The arrow marks in panel d refer to the decomposition products that form in preference to LiFePO$_4$ co-existing with the aqueous solution at different pH and applied potentials.}
\label{fig:2}
\end{figure}

\noindent We summarize the querying of structural and thermodynamic data for Li, Na, K, Mg, and Ca based materials from MP (v2025.06.09), in panels a and b of \textbf{Figure~{\ref{fig:2}}}. Since intercalation electrodes contain at least one redox-active transition metal (TM) and one anion to maintain charge neutrality with (de)intercalation, we searched for solids containing at least one and at most three TMs and at least one and at most three anions. We chose at most three TMs and anions to keep the number of possible elemental combinations computationally tractable. Also, we required that the solids contain at least one of the electroactive ions among Li, Na, K, Mg, and Ca, reflecting the given structure's ability to host one of the cations considered. Note that we excluded recent mixing corrections on total energies \cite{kingsbury2022flexible} that have been made available for a small fraction of materials in the MP to maintain energy (and voltage) consistency across the diverse chemical space screened here.

For the choice of TMs, we included all 3$d$ metals that show redox activity (i.e., from Ti to Cu), and selected 4$d$ and 5$d$ metals that have shown redox activity and have been part of battery cathode explorations, as highlighted in blue in \textbf{Figure~{\ref{fig:2}}a} \cite{yabuuchi2015high, wang2021tuning, ichitsubo2014new, sathiya2013reversible, wu2022highly, pearce2017evidence, bashian2018correlated}. All anions considered are shown in yellow in \textbf{Figure~{\ref{fig:2}}a}. Our search lead us to 12,536 Li, 3,598 Mg, 1,344 Ca, 1,783 Na, and 1,153 K containing structures, resulting in a total of 20,414 possible electrode materials, as shown by the cyan bars in \textbf{Figure~\ref{fig:2}b}. The disproportionately high number of Li-containing structures in \textbf{Figure~\ref{fig:2}b} is a consequence of the heavily investigated LIB cathode chemical space and is not necessarily an indication that there are intrinsically more Li intercalating structures.

\subsection{Constructing seawater Pourbaix diagrams}
\label{sec:seawater-Pourbaix}
\noindent While Pourbaix diagrams involving elemental systems are well documented \cite{pourbaix1966atlas}, our queried electrodes contain multiple elements, requiring the construction of challenging multi-element Pourbaix diagrams \cite{thompson2011pourbaix}. Hence, we use the thermodynamic framework \cite{persson2012prediction, patel2019efficient, sun2019non, singh2017electrochemical} encoded in the ``\texttt{pourbaix\_diagram}" module of the \texttt{pymatgen} \cite{ong2013python} package to generate the required multi-element Pourbaix diagrams. The \texttt{pourbaix\_diagram} module combines experimental ionic solubility data \cite{johnson1992supcrt92, pourbaix1974atlas} and computational solid phase data, as documented in MP, and identifies phases that minimize the semi-grand canonical potential that is open to H and O (also referred to as the `Pourbaix potential'). In all, we constructed 7,806 Pourbaix diagrams spanning over 1,046 distinct multi-element systems, based on our queried MP data. Mathematical details of the Pourbaix potential and the associated thermodynamic formalism can be found in \textbf{Section~S2}.

Example Pourbaix diagrams involving olivine-LiFePO$_4$ are shown in panels c and d of \textbf{Figure~{\ref{fig:2}}}, indicating the regions of pH and applied potentials (in V vs.~SHE) where the phase of interest is thermodynamically stable (for example, the purple region in \textbf{Figure~{\ref{fig:2}}c}). Note that water is stable only over a limited window in any Pourbaix diagram, as indicated by the region within the dashed red lines in \textbf{Figure~{\ref{fig:2}}c} and \textbf{d}, with the red lines corresponding to hydrogen and oxygen evolution reactions. Thus, an intersection of a solid's region of stability and the water stability window indicates that the solid is thermodynamically stable in water, over the intersected region.

The LiFePO$_4$ Pourbaix diagrams also illustrate the effect of the composition of the aqueous medium on the solid's aqueous stability. For example, in an aqueous solution containing large concentrations of ions, namely, [Li] = 0.75~M, [Fe] = 0.25~M, and [P] = 0.33~M, we observe LiFePO$_4$ to be thermodynamically stable (purple region in \textbf{Figure~{\ref{fig:2}}c}) across a range of pH and applied potentials. However, in a dilute solution containing [Li] = [Fe] = [P] = $10^{-6}$~M, LiFePO$_4$ does not appear as a stable phase on the Pourbaix diagram at any pH or applied potential, indicating its instability. Indeed, across a subset of the pH and applied potential region where LiFePO$_4$ is stable in \textbf{Figure~{\ref{fig:2}}c}, we observe different Fe and Li-containing phases to coexist in preference to LiFePO$_4$ (each distinct colour in \textbf{Figure~{\ref{fig:2}}d}, corresponds to different combination of stable phases). Thus, our observation with LiFePO$_4$ clearly shows, perhaps not surprisingly, that the stability of a solid is influenced by its surrounding aqueous (solution) composition.

We fixed the concentration of ions in the aqueous medium to be the same as the average seawater concentration in our Pourbaix diagram constructions, given that many brines are derived from seawater. Note that while seawater contains numerous elements, the six most abundant species are Cl$^{-}$, Na$^{+}$ , Mg$^{+2}$, SO$_4^{-2}$, Ca$^{+2}$, and K$^{+}$, as shown in \textbf{Figure~S2a} of the SI, while all other elements/species occur in low concentrations ($<$0.01~M). Since the computational complexity of a Pourbaix diagram increases significantly with every additional element (see \textbf{Figure~S2b}) \cite{patel2019efficient}, we only added the top four abundant seawater elements, i.e., Cl, Na, Mg and S, to the solution phase in addition to the elements present in the solid. The resultant Pourbaix diagrams that are generated are our seawater Pourbaix diagrams, which we subsequently use to estimate the (meta)stability of the queried solid phases. More details on the construction of the seawater Pourbaix diagrams, including modifications made to the Pourbaix potential, are documented in \textbf{Section~S3} of the SI. While we consider the average seawater concentration in our work, our framework can be trivially updated to consider any other desired aqueous composition.

\subsection{Determining aqueous (meta)stability}
\label{sec:metastability}
\begin{figure}[h!]
\centering
\includegraphics[width=\textwidth]{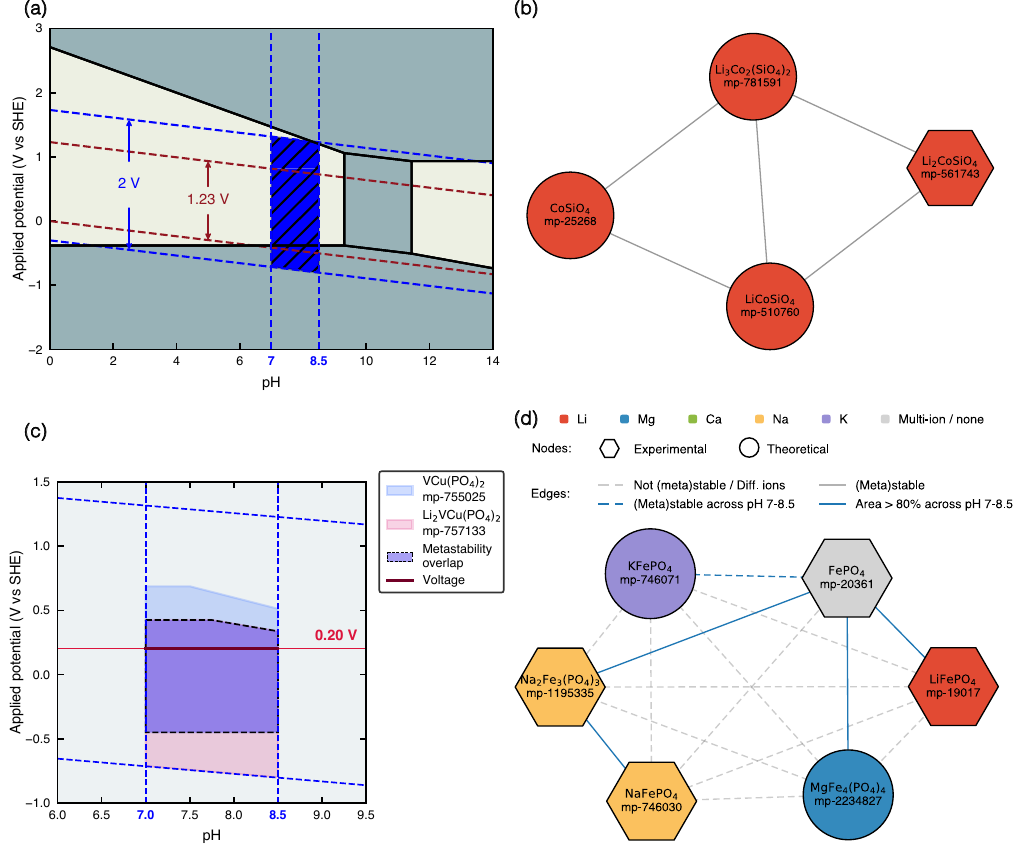}
\caption{(a) Sample Pourbaix diagram showing the seawater stability window in hashed blue. (b) Sample computational graph generated for the CoSiO$_4$ host. The text annotations provide the MP ID for each intercalated structure constituting the graph. (c) Zoomed in seawater stability window (area within the dashed blue lines) showing the intersection of the metastability areas of an intercalated-deintercalated (Li$_2$VCu(PO$_4$)$_2$-VCu(PO$_4$)$_2$) pair of structures and their associated operating voltage of Li (de)intercalation (solid red line). (d) A sample computational graph of olivine-FePO$_4$ host, which unlike panel (b), encodes information across all the five electroactive ions considered. Since olivine-FePO$_4$ can intercalate four different cations, the graph contains nodes of four colors, namely, red (containing or exchanging Li$^+$), blue (Mg$^{+2}$), yellow(Na$^+$) and purple(K$^+$). The gray colored FePO$_4$ node represents a fully deintercalated structure that can intercalate multiple ions. The graph also encompasses information about the metastability/co-existence of pairs of structures via distinct edges.}
\label{fig:3}
\end{figure}

\noindent Since seawater exhibits an average pH of 8 \cite{jiang2019surface}, we don't expect seawater to be stable under highly basic (e.g., pH~=12) or highly acidic (e.g., pH~=~2) conditions, and hence set aside a pH range of 7$-$8.5 as the region of stability of seawater. Note that we use seawater as a proxy in our work, but our method can be employed for any solvent conditions encountered in geothermal brines. Additionally, we shift the hydrogen evolution line by $-$0.3~V vs.~SHE and the oxygen evolution line by $+$0.5~V vs. SHE to account for overpotentials for the hydrogen and oxygen evolution reactions \cite{suo2015water}, resulting in the expansion of the water stability window from 1.23~V to $\sim$2~V. Thus, we refer the `seawater stability window' as the combination of the applied potential range of $\sim$2~V and a pH range of 7$-$8.5, as highlighted by the hashed blue area in \textbf{Figure~{\ref{fig:3}}a}. To determine the seawater stability of each of the queried structures, we examined the overlap between the seawater stability window, and the stability region of the electrode, where an overlap indicates that the electrode can be stable in seawater. 

Upon computing the seawater stability for all the queried electrodes, we find only 6 Li, 4 Mg, 3 Ca, 6 Na, and 7 K structures to be stable in seawater, as indicated by the yellow bars in \textbf{Figure~{\ref{fig:2}}b}. The list of seawater stable electrodes, which have a stable domain within the seawater-stability window are compiled in \textbf{Table~S2} of the SI. Surprisingly, the widely used Li-recovery electrodes that have been demonstrated to work with seawater-like brines, olivine-LiFePO$_4$ and spinel-LiMn$_2$O$_4$ \cite{pasta2012batteries, liu2020lithium, kim2018electrochemical, trocoli2015nickel, trocoli2017optimized, zhao2019highly}, are not among the list of stable electrodes in \textbf{Figure~{\ref{fig:2}}b}. This suggests that both LiFePO$_4$ and LiMn$_2$O$_4$ could be metastable in seawater and highlights the need to evaluate and quantify the extent of `reasonable' metastability that allows electrodes to function in seawater. 

Previous work has shown that solid phases can be assumed to be metastable in water if at least one of the stable decomposition products in the aqueous medium is a solid, and the energy of decomposition is $\leq$0.5~eV~atom$^{-1} $\cite{singh2017electrochemical}. The basis for this metastability criterion is that solid-solid transitions typically exhibit  high activation barriers, indicating that decomposition processes with `small' thermodynamic driving forces ($\leq$0.5~eV~atom$^{-1}$) may not be kinetically active. Additionally, a solid decomposition product, if formed, can create a protective coating over the original decomposing solid to prevent further decomposition. While this metastability criterion was developed for phases in dilute aqueous solutions, we modified and extended this criterion to our seawater Pourbaix diagrams (see \textbf{Sections~S5} and \textbf{S6} for details).

To modify the metastability criterion, we estimated the decomposition energy of olivine-LiFePO$_4$ in seawater, which is $\leq$0.28~eV~atom$^{-1}$ across the seawater stability window, with an Fe-containing solid decomposition product (see \textbf{Figure~S4a}). Since olivine-LiFePO$_4$ has shown the ability to recover Li from brines \cite{pasta2012batteries, liu2020lithium, trocoli2015nickel} and we would want any new electrode material to be at least as (meta)stable as LiFePO$_4$, we selected a decomposition energy threshold of 0.3~eV~atom$^{-1}$ instead of the previously used 0.5~eV~atom$^{-1}$ threshold \cite{singh2017electrochemical}. Additionally, it is desirable for the solid decomposition product that forms to be in the deintercalated state (i.e., without containing Li in the case of LiFePO$_4$) so that it does not trap any of the working ions from the intercalation electrode. Also, it is undesirable for the decomposed solid to be an element (e.g., LiFePO$_4$ decomposing to form Fe), since pure elements, especially metals, are unlikely to form protective layers that are permeable to the working ion. Thus, our modified metastability criterion for any identified electrode is that the structure should form a non-elemental solid decomposition product with a maximum decomposition energy of 0.3~eV~atom$^{-1}$ within the seawater stability window (see \textbf{Section~S6}). 

Note that Pourbaix diagrams are constructed using non-H and non-O elemental ratios that are chosen to match the elemental ratios of the solid under consideration \cite{garay2026enriching}. For example, to evaluate the aqueous stability of solid LiFePO$_4$, the elemental ratios of the corresponding Pourbaix diagram has to be Li:Fe:P~=~1:1:1. Thus, as long as the elemental ratios of the Pourbaix diagram matches the solid of interest, the decomposition energy at any pH and potential is simply the difference between the solid's Pourbaix potential and the sum of the Pourbaix potentials of the corresponding decomposition products. However, in our seawater Pourbaix diagrams, the elemental ratios of the decomposing solid do not always match those of the Pourbaix diagram due to the presence of additional elements in seawater (e.g., Cl) in non-dilute limits. Thus, to evaluate the decomposition energy of a queried electrode in a seawater Pourbaix diagram (see \textbf{Section~S5}), we $i$) identify possible species that can react with the solid to form the decomposition products at each pH and potential, $ii$) enumerate all possible reactions, and $iii$) identify the reaction that yields the highest decomposition energy (i.e., maximizing the thermodynamic driving force). We subsequently compare the highest decomposition energy with the 0.3~eV~atom$^{-1}$ threshold. 

Using our modified metastability criterion, we found 10,131 Li, 2,437 Mg, 1,344 Ca, 1,185 Na, and 559 K phases to be metastable in seawater, as indicated by the violet bars in \textbf{Figure~{\ref{fig:2}}b}. As expected and per our definition, our modified criterion identifies both olivine-LiFePO$_4$ and spinel-LiMn$_2$O$_4$ as metastable in seawater, in agreement with experimental observations so far. Once we identify a structure to be metastable, we can project the region within the seawater stability window of the corresponding Pourbaix diagram where the structure is metastable. As an example, we have shown the metastability regions of VCu(PO$_4$)$_2$ (blue region) and Li$_2$VCu(PO$_4$)$_2$ (pink region) in the magnified version of the seawater Pourbaix diagram in \textbf{Figure~{\ref{fig:3}}c}.

\subsection{Estimating (de)intercalation voltages and metastability overlap}
\label{sec:voltages}
\noindent The materials that have been shortlisted as metastable could $-$ in principle $-$ be metastable across any region within the seawater stability window. Thus it is important to examine whether the regions of metastability overlap for an electroactive ion intercalated structure as well as its corresponding deintercalated host, similar to the overlap seen (dark purple region) in the case of Li$_2$VCu(PO$_4$)$_2$-VCu(PO$_4$)$_2$ in \textbf{Figure~{\ref{fig:3}}c}. Equally important is to ascertain whether the queried phases are metastable at their corresponding ion intercalation voltage, i.e., do the structures remain metastable as they exchange electroactive ion(s). To address these aspects, we computed the average topotactic intercalation voltages of all the shortlisted frameworks using the Nernst equation and based on DFT-calculated total energies from MP \cite{aydinol1997ab}. 

To determine topotactic average voltages, we need to identify intercalated-deintercalated structure `pairs' that are `similar` to one another, since topotactic (de)intercalation occurs with minimal changes in the `host' (host is the structure without the electroactive ion). While some electrodes have (meta)stable hosts already available in MP, like FePO$_4$ in olivine LiFePO$_4$, other frameworks, such as the double-perovskite Ca$_2$FeMoO$_6$, have an unstable host (FeMoO$_6$) that is absent from the MP database. Thus, if a (meta)stable host exists, it can possibly co-exist with an intercalated state of the electrode at a given average voltage, such as FePO$_4$ co-existing with LiFePO$_4$ at an average voltage of $\sim$0.46~V vs. SHE (see \textbf{Figure~S5a}) \cite{malik2011kinetics}. Therefore, to evaluate intercalation voltages across materials, we query the fully deintercalated structure for each seawater (meta)stable intercalated structure from MP and subsequently evaluate the seawater metastability of the host. Note that a given intercalated structure can be identified as being similar to several deintercalated structures (including structures with minor distortions and partial deintercalation) and vice-versa, implying the need to identify common host frameworks across multiple structures (see \textbf{Section~{\ref{sec:comparison}}} for details).

After identifying all the topotactic (de)intercalated states of an electrode host, we determine the states that can thermodynamically co-exist at a constant voltage using pseudo-binary convex hulls, where the composition of the electroactive ion is varied across the (de)intercalated structures. Using DFT-calculated energies from the MP, we construct a 0~K pseudo-binary hull for each combination of a host and an electroactive ion. Note that the pseudo-binary hulls are used solely to determine possible coexisting phases, not to reassess the seawater (meta)stability (see \textbf{Figure~S6} for a schematic and further explanation).

Specifically, only phases connected by a facet (or a tie-line) of the pseudo-binary convex hull are thermodynamically allowed to coexist and we compute the average intercalation voltage (versus SHE) for each pair of coexisting phases. If the calculated voltage lies within the seawater metastability window, we check the metastability of both the coexisting phases and screen for pairs that remain metastable across the entire 7-8.5 seawater pH window at their intercalation voltage (see the Li$_2$VCu(PO$_4$)$_2$-VCu(PO$_4$)$_2$ example in \textbf{Figure~{\ref{fig:3}}c}). If the seawater metastability areas of two phases on a facet do not overlap or if the calculated voltage of a coexisting pair falls outside the overlapping region of metastability of the individual phases (see example in \textbf{Figure~S6c1, c2}), the corresponding intercalated phase is removed from the pseudo-binary hull and the hull is recomputed. We repeat this procedure recursively until all suitable candidate pairs, i.e., intercalated-deintercalated structure pairs with overlapped metastability region and voltage within the seawater stability window, of a given host are identified or all structures mapped to a host are exhausted. 

It is important to note that under operating electrochemical conditions, operating voltages for electrodes can deviate away from our predicted values due to overpotentials as well as any phase transitions in the electrode material. Hence, it is necessary that a candidate electrode is stable over a wider range of potentials around the computed voltage. For example, olivine-LiFePO$_4$ exhibits an average voltage of 0.46~V vs.~SHE, which is approximately equidistant from both the H and O evolution lines within the seawater stability window (\textbf{Figure~S5a}). Additionally, the metastability regions of olivine-LiFePO$_4$ and its deintercalated-FePO$_4$ states overlap over 88\% of the seawater metastability window (\textbf{Figure~S5a}), enabling the coexisting phases to accommodate large overpotentials. Thus, the percentage overlap of an electrode's metastability region with the seawater metastability window is an important metric to quantify the utility of an electrode under operating conditions. 

We use a percentage metastability overlap metric of $\geq$80\%, since this is similar to the overlap exhibited by olivine-LiFePO$_4$ ($\sim$88\%, \textbf{Figure~S5a}), to screen for electrode candidates that are metastable and exhibit a voltage within the seawater stability window. Note that our metastability overlap metric requires the area of intersection of the metastable regions of both the intercalated and deintercalated phases to constitute $\geq$80\% of the seawater stability window. For instance, the Li$_2$VCu(PO$_4$)$_2$-VCu(PO$_4$)$_2$ pair in \textbf{Figure~{\ref{fig:3}}c} exhibits an overlap of the metastability regions of the individual phases and an average intercalation voltage of $\sim$0.20~V vs.~SHE that is within the seawater stability window. However, the pair's metastability overlap constitutes $<$80\% of the overall seawater stability window and hence is not considered a candidate for further analysis. 

In summary, in this step, we screen for pairs of deintercalated-intercalated structures with $i$) an intersecting (or overlapping) region of metastability, $ii$) an average intercalation voltage that lies within the intersecting metastability region across the entire 7-8.5 pH range, and $iii$) the intersecting metastability area occupying at least 80\% of the seawater stability window.

\subsection{Comparing framework structures}
\label{sec:comparison} 
\noindent To map and visualize all unique intercalated-deintercalated combinations with similar hosts for voltage and metastability overlap calculations (\textbf{Section~{\ref{sec:voltages}}}), we constructed computational graphs, as shown in \textbf{Figure~{\ref{fig:3}b}} and \textbf{d}. Circular and hexagonal nodes in our graphs represent theoretical and experimental structures, respectively, with the color of a node representing the electroactive ion associated with the host. Gray colored nodes typically reflect deintercalated hosts that can accommodate multiple types of cations. Edges in our graphs connect structures with similar underlying hosts, with the type of the edge (solid or dashed and blue or gray) encoding information on the metastability of the connected nodes (see \textbf{Figure~{\ref{fig:3}}d}). We use \texttt{pymatgen}'s \texttt{StructureMatcher} module, with its default tolerances, to identify whether two underlying hosts are similar or not. Thus, each graph is a collection of one or more interconnected nodes, with each node representing various intercalated states of a host and every pair of connected nodes signifying a possible topotactic intercalated-deintercalated combination. 

As an example, \textbf{Figure~{\ref{fig:3}}b} shows the graph for a polymorph of CoSiO$_4$ host (with a MP ID of mp-25268), comprising four compositions $-$ CoSiO$_4$, LiCoSiO$_4$, Li$_2$CoSiO$_4$, and Li$_3$CoSiO$_4$. The edges among the CoSiO$_4$, LiCoSiO$_4$, and Li$_3$CoSiO$_4$ nodes indicate that these three structures share similar underlying CoSiO$_4$ hosts (i.e., the structures are similar if all Li are removed from each structure). Notably, the Li$_2$CoSiO$_4$ node connects to LiCoSiO$_4$ and Li$_3$CoSiO$_4$ but not to CoSiO$_4$, which can be attributed to deviations in the CoSiO$_4$ host structure associated with Li$_2$CoSiO$_4$ that does not match well with the pure CoSiO$_4$ host, per the default tolerances of the \texttt{StructureMatcher} module. We interpret this to mean that Li insertion into CoSiO$_4$ generally distorts the pure CoSiO$_4$ host, with the distortion at Li$_2$CoSiO$_4$ being particularly strong. Although every node in this graph (\textbf{Figure~{\ref{fig:3}}b}) is not connected to every other node, their presence in a single graph demonstrates that they share a common host, with distortions arising from the (de)intercalation of the electroactive ion. Thus, we classify all structures being part of a graph to exhibit a similar underlying host, as long as a given structure shares at least one edge to another structure in the graph. 

Upon identifying electrodes with suitable operating voltages and high metastability area overlap in \textbf{Section~{\ref{sec:voltages}}}, we compare the hosts across different working ions to distinguish the structures that intercalate a single ion (for ion recovery) from those that can intercalate multiple ions (for desalination). Specifically, we construct computational graphs for all structures across the five electroactive ions, as shown for the example of olivine-FePO$_4$ in \textbf{Figure~{\ref{fig:3}d}}. Physically, each graph (that contain a cluster of interconnected nodes) corresponds to a single host structure that is capable of intercalating one or multiple cations, thus encoding the ion-selectivity information. Importantly, solid blue edges in our graphs (\textbf{Figure~{\ref{fig:3}}d}) indicate that the two nodes are not only seawater metastable at their (de)intercalation voltage but also exhibit an interesecting metastability area that occupies $\geq$80\% of the seawater stability window. Thus, we enlist all nodes (i.e., structures) that are connected by solid blue edges as the most promising candidates for ion capture and include the information on whether a given host is selective to a specific cation or it can intercalate multiple cations as well.

\section{Results}
\subsection{Screening statistics}
\begin{figure}[h!]
\centering
\includegraphics[width=\textwidth]{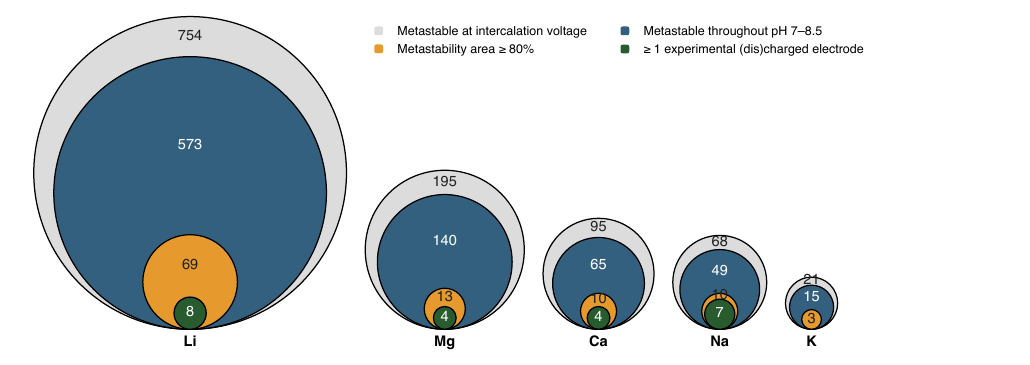}
\caption{Venn diagrams showing the reduction in the number of candidate electrodes with sequential screening. Gray circles indicate the number of intercalation-deintercalation pairs that are metastable at their operating voltage, while blue circles enumerate pairs that are metastable across the entire 7-8.5 seawater pH window. Yellow circles indicate the number of pairs which exhibit an intersecting metastability area that occupies more than 80\% of the seawater stability window. Green circles signify the subset of pairs that have at least one experimentally known structure.}
\label{fig:4}
\end{figure}

\noindent In total, we evaluate the seawater metastability of 12,536 Li, 3,598 Mg, 1,344 Ca, 1,783 Na and 1,153 K intercalation electrodes, as queried from MP (cyan bars in \textbf{Figure~{\ref{fig:2}}b}), with most of the queried electrodes, namely 10,131 Li, 2,437 Mg, 780 Ca, 1,185 Na and 559 K, being metastable in seawater at some pH and applied potential (purple bars). Subsequently, screening for intercalation-deintercalation pairs that are metastable at their intercalation voltage reduces the number of candidates significantly, yielding possibly 754 Li, 195 Mg, 95 Ca, 68 Na, and 21 K hosts, as summarized by the gray circles in the Venn diagrams of \textbf{Figure~{\ref{fig:4}}}. Application of the screening criteria, as developed in \textbf{Sections~{\ref{sec:voltages}}} and \textbf{{\ref{sec:comparison}}}, causes further reductions in the number of possible electrodes. For instance, requiring the intercalation-deintercalation pairs to be metastable at their intercalation voltage, across the entire 7-8.5 seawater pH range, and ensuring an intersecting area that occupies at least 80\% of the seawater stability window (as outlined in \textbf{Section~\ref{sec:voltages}}) results in a total of only 69 Li, 13 Mg, 10 Ca, 10 Na, and 3 K hosts, displayed as yellow circles in \textbf{Figure~{\ref{fig:4}}}. Thus, we deem all the hosts constituting the yellow circles in \textbf{Figure~{\ref{fig:4}}} to be potential candidates for electrochemical ion capture. Finally, there are 8 Li, 4 Mg, 4 Ca, 7 Na, and 0 K hosts that have at least one experimental phase among the intercalated-deintercalated pairs (green circles in \textbf{Figure~{\ref{fig:4}}}), highlighting the strongest set of candidates for experimental trials of ion capture. The final list of experimentally known candidates, alongside the most promising theoretical candidates are compiled in \textbf{Table~{\ref{table:1}}} and in the sections below. 

\subsection{Ion selectivity statistics}
\noindent The Venn diagrams in \textbf{Figure~{\ref{fig:4}}} do not show the intersection of the sets across the electroactive ions considered, i.e., the statistics represent those for a single electroactive ion and do not include ionic selectivity. Therefore, all structures within the gray sets of \textbf{Figure~{\ref{fig:4}}} are nodes in our computational graphs (see the example in \textbf{Figure~{\ref{fig:3}}d}). The connectivity of edges in a graph encodes the ion-selectivity information, thus motivating our use of computational graphs in this work. Another reason for the use of graphs is to overcome the limitations of finite and default tolerances that are often used in matching structures/hosts, since there are no universal set of tolerances that work well across different structural groups and their types of distortions. Importantly, we compile the cases of hosts belonging to the gray sets of \textbf{Figure~{\ref{fig:4}}} that can intercalate single or multiple cations via our graph constructions and depict the statistics in \textbf{Figure~{\ref{fig:5}}}.

Graphs (or hosts) that can intercalate only one cation are depicted as having nodes of the same color, namely, red, blue, green, yellow, and violet colors for Li, Mg, Ca, Na, and K, respectively, in \textbf{Figure~{\ref{fig:5}}a}. Most of the hosts that intercalate only a single cation are Li-selective (587 hosts, \textbf{Figure~{\ref{fig:5}}a}), reflecting the larger volume of available Li data, followed by 94 Mg, 27 Ca, 11 Na, and 8 K selective hosts. Graphs displaying nodes of multiple colors, as shown in panels b, c, and d of \textbf{Figure~{\ref{fig:5}}}, represent hosts that can intercalate more than one cation. Among such hosts, we identify 115 frameworks that intercalate two cations (\textbf{Figure~{\ref{fig:5}}c}), 28 hosts that intercalate three cations (\textbf{Figure~{\ref{fig:5}}d}), and five hosts that intercalate  four cations (\textbf{Figure~{\ref{fig:5}}b}). Notably, our screening identifies no host structure that can intercalate all five electroactive cations considered in this work, indicating that there is no perfect electrode candidate for desalination.

\subsection{Ion-selective and non-selective electrodes}
\begin{figure}[h!]
\centering
\includegraphics[width=\textwidth]{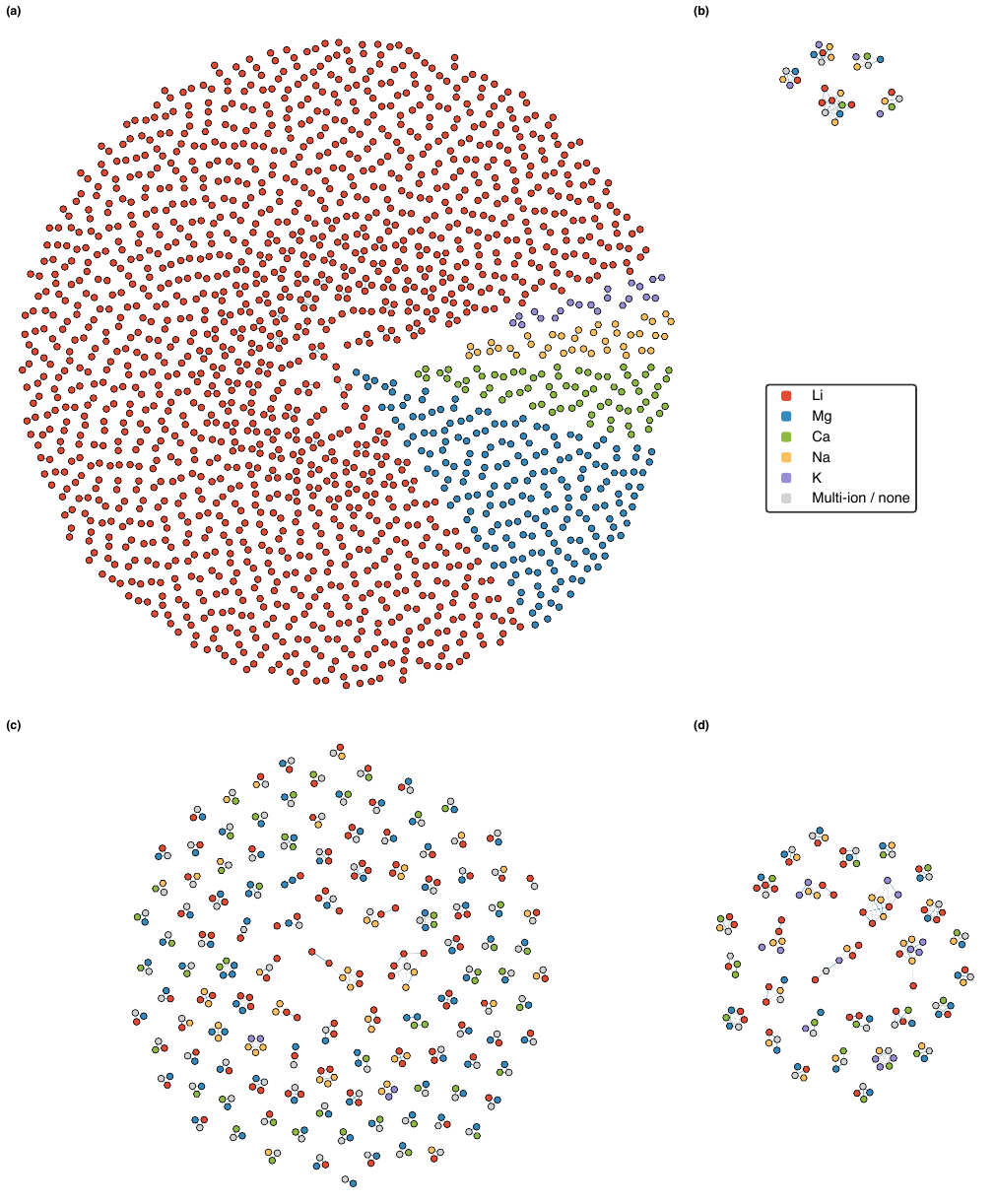}
\caption{Graphs of all hosts that can intercalate (a) one, (b) four, (c) two, and (d) three distinct cations, corresponding to a total of 859, 5, 115, and 28 hosts, respectively.}
\label{fig:5}
\end{figure}

\noindent While the collection of graphs in \textbf{Figure~{\ref{fig:5}}} show numerous ion-selective/non-selective hosts, we are interested in the most promising among them, as represented by the frameworks in the yellow or green sets of \textbf{Figure~{\ref{fig:4}}}. Hence, we compile the set of candidates that we identify as most promising across all electroactive cations and for both ion-selective and non-selective use cases in \textbf{Figure~{\ref{fig:6}}}. The most deintercalated state of the host (and its MP ID) is labeled along the horizontal axis for the case of ion-selective (panel a) and non-selective (panel b) scenarios in \textbf{Figure~{\ref{fig:6}}}, with the vertical axis in both panels indicating the intercalation voltage. Note that each voltage value corresponds to the coexistence of an intercalated and a deintercalated phase, with circular and hexagonal shapes indicating cases with only theoretical structures and at least one experimental structure in a given pair, respectively. Color coding in \textbf{Figure~{\ref{fig:6}}a} and \textbf{b} is identical to \textbf{Figure~{\ref{fig:5}}}, with the horizontal blue dashed lines representing the narrowest voltage window of seawater stability across the 7-8.5 pH range.

Among the screened metastable hosts constituting the yellow sets of \textbf{Figure~{\ref{fig:4}}}, 66 Li, four Mg, three Ca, three Na, and two K hosts intercalate only a single ion and qualify as ion-selective electrodes. Given the large number of Li-selective electrodes in the yellow set, we considered the Li-selective electrodes in the experimental (green) subset to be the most promising, which include LiFeP$_2$O$_7$ (mp-19294, \textbf{Figure~{\ref{fig:6}}c}), Fe$_2$(PO$_4$)$_3$ (mp-26412, \textbf{Figure~{\ref{fig:6}}d}) and Fe$_2$(MoO$_4$)$_3$ (mp-21925, \textbf{Figure~{\ref{fig:6}}e}). Among the screened ion-selective hosts for Mg, Ca, and Na, we identify two hosts in the green subset for each cation, namely, NbCoO$_4$ (mp-31518), and TiFe$_2$O$_5$ (mp-35482) for Mg, Ca$_3$Co$_2$(Si$_2$O$_7$)$_2$ (mp-1256217), and CaFeMoO$_6$ (mp-1368738) for Ca, and Fe$_7$(PO$_4$)$_6$ (mp-510351), and NaFe$_3$(PO$_4$)$_3$ (mp-556692) for Na. Notably, all candidates identified as promising K-selective are theoretical hosts.

\textbf{Figure~{\ref{fig:6}}b} compiles the promising candidate hosts (i.e., from the yellow sets of \textbf{Figure~{\ref{fig:4}}}) for intercalating two cations (blue shaded region), three cations (pink shaded), and four cations (green shaded) simultaneously, thus representing the shortlisted candidates for non-selective ion capture. For the case of hosts that can capture Li and other cation(s), we prioritized the listing of experimentally known hosts in \textbf{Figure~{\ref{fig:6}}b}, similar to \textbf{Figure~{\ref{fig:6}}a}. Faded symbols in \textbf{Figure~{\ref{fig:6}}b} indicate that the intercalated-deintercalated pair involving the specific electroactive ion is not seawater metastable during operation, while the hatched symbols indicate pairs that have an intersection area that occupies 70\% to 80\% of the seawater stability window across the full 7-8.5 pH range. In the case of non-selective hosts, we do obtain frameworks that are not as metastable for specific electroactive ions during our screening process (i.e., the faded and hatched symbols) since all structures across different electroactive ions that are identified as similar belong to the same graph. While we don't expect hosts indicated by faded symbols to be promising for the specific cation, hosts highlighted by hatched symbols may still be relevant for experimental trials involving stricter potential controls.

Importantly, several hosts in \textbf{Figure~{\ref{fig:6}b}} show a `small' difference ($<$0.1~V) in the intercalation voltage among different cations and are hence likely to intercalate multiple ions at the same overpotential experimentally. Examples include the different polymorphs of FePO$_4$ (mp-25303,  mp-752872, and mp-776288) that intercalate two cations, the Ca-Mg intercalating Fe$_2$(MoO$_4$)$_3$ (mp-1042928), the Na-Li intercalation in Fe(PO$_3$)$_3$ (mp-25310) and NaCoPCO$_7$ (mp-1101670), the Na-K intercalation in LiFePCO$_7$ (mp-1176697), and Mg-K intercalation in olivine-FePO$_4$ (mp-20361). Other hosts in \textbf{Figure~{\ref{fig:6}}b} exhibit voltage differences that are $>$0.1~V among the intercalating cations, indicating that these electrodes may be better used in selectively intercalating a specific cation using potentiostatic conditions.

\subsection{Shortlisted ion-selective hosts}
\label{subsec: Shortlisted ion-selective hosts}
\begin{figure}[h!]
\centering
\includegraphics[width=\textwidth]{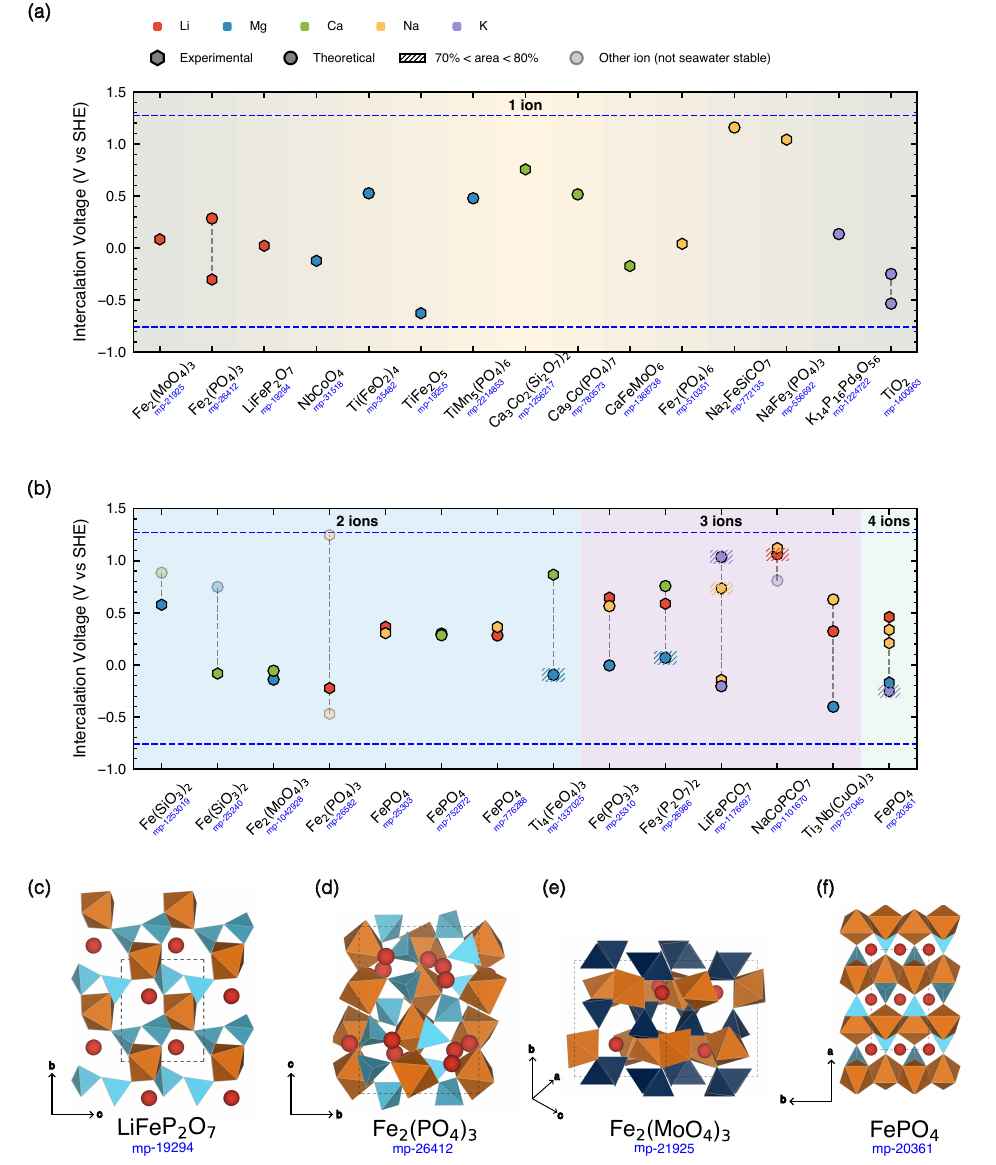}
\caption{Intercalation voltages of the shortlisted (a) ion-selective and (b)  non-selective intercalating hosts. (c)-(e) Structures of promising Li-selective electrodes where the light blue tetrahedra, orange octahedra, and dark blue tetrahedra indicate PO$_4$, FeO$_6$, and MoO$_4$, respectively. (f) Olivine-FePO$_4$ with the same colour scheme as panels (c)-(d). Red spheres in panels (c)-(e) show sites occupied by electroactive cations in the lattice.}
\label{fig:6}
\end{figure}

\noindent In this section, we provide detailed descriptions of the host structures that are identified as most promising ion-selective hosts for each electroactive ion considered and compiled in \textbf{Figure~{\ref{fig:6}}a}. For non-Li intercalating hosts, we provide descriptions of theoretical hosts that we have identified with our workflow, which are potential candidates as well. The constructed seawater Pourbaix diagrams for all candidates described here are compiled in \textbf{Figure~S7} of the SI, while the corresponding computational graphs are shown in \textbf{Figure~S11}.

\subsubsection{Li-selective}
\noindent LiFeP$_2$O$_7$ (mp-19294, \textbf{Figure~{\ref{fig:6}}c}) belongs to the pyrophosphate \cite{hautier2011phosphates} family of compounds with a monoclinic $P2_1$ spacegroup, tetrahedrally coordinated PO$_4$ and octahedrally coordinated FeO$_6$. Each PO$_4$ tetrahedron shares one corner with another PO$_4$ tetrahedron, forming P$_2$O$_7$ units. While each PO$_4$ shares three corners with three different FeO$_6$ octahedra, each FeO$_6$ is connected to five different P$_2$O$_7$ groups, where one of the P$_2$O$_7$ groups acts as a chelating sequence around the Fe \cite{wurm2002lithium, padhi1997effect}, resulting in FeP$_2$O$_7$ units that create open channels for Li intercalation. LiFeP$_2$O$_7$ forms as the fully deintercalated phase experimentally with Fe in the +3 oxidation state. Li intercalation reduces the Fe to +2 and yields Li$_{1+x}$FeP$_2$O$_7$, which has been experimentally reported for $0 \leq x \leq 0.8$ \cite{wurm2002lithium}. However, at $x = 1$, the topotactic $P2_1$ intercalated phase (Li$_2$FeP$_2$O$_7$, mp-779056) remains a theoretical structure, albeit with the intersecting metastability area occuping 88\% of the seawater stability window (\textbf{Figure~S7d}). Our estimated intercalation voltage (0.02~V vs.~SHE or 3.06~V vs.~Li) for the LiFeP$_2$O$_7$---Li$_{2}$FeP$_2$O$_7$ pair is in close agreement with the experimentally reported two-phase plateau of 2.92~V vs.~Li \cite{wurm2002lithium, padhi1997effect}. 

Li$_3$Fe$_2$(PO$_4$)$_3$ (mp-25993) has a monoclinic anti-NASICON (Na superionic conductor) structure in the $P2_1/c$ spacegroup, consisting of PO$_4$ tetrahedra and two crystallographically distinct FeO$_6$ octahedra. Each FeO$_6$ octahedra shares all its edges with PO$_4$ tetrahedra and vice versa, leading to a three dimensional framework allowing for three distinct Li sites with different coordination environments. Li$_3$Fe$_2$(PO$_4$)$_3$ has Fe in its +3 oxidation state and experiments report two distinct voltage plateaus upon intercalation, resulting in Li$_4$Fe$_2$(PO$_4$)$_3$ and Li$_5$Fe$_2$(PO$_4$)$_3$ \cite{patoux2003comparative}. Note that Li$_3$Fe$_2$(PO$_4$)$_3$ shares the same host lattice as Fe$_2$(PO$_4$)$_3$ (mp-26412, \textbf{Figure~{\ref{fig:6}}d}), which is the most deintercalated composition in the graph and is available in MP as well, explaining its listing in \textbf{Figure~{\ref{fig:6}}a}. However, Fe$_2$(PO$_4$)$_3$ is not seawater stable and only Li$_3$Fe$_2$(PO$_4$)$_3$ - Li$_5$Fe$_2$(PO$_4$)$_3$ and Li$_2$Fe$_2$(PO$_4$)$_3$ - Li$_5$Fe$_2$(PO$_4$)$_3$ pairs are metastable with an intersecting area of over 80\% of the seawater stability window (see \textbf{Figure~S7b1,b2}). Here, we consider the Li$_3$Fe$_2$(PO$_4$)$_3$ - Li$_5$Fe$_2$(PO$_4$)$_3$ pair, with a voltage of $-0.3$~V vs.~SHE, to be the most promising since this pair has been experimentally reported \cite{patoux2003comparative}. Also, Li$_3$Fe$_2$(PO$_4$)$_3$ and FeP$_2$O$7$ have been explored as an intercalation electrodes in aqueous LIBs \cite{paharik2017iron}. 

Fe$_2$(MoO$_4$)$_3$ (mp-21925, \textbf{Figure~{\ref{fig:6}}e}) exhibits an orthorhombic $Pbcn$ spacegroup containing FeO$_6$ octahedra and MoO$_4$ tetrahedra that share their corners with distinct MoO$_4$ and FeO$_6$ groups, respectively. As a result, the framework forms intercalation sites that are interconnected in all three dimensions. Topotactic lithiation of Fe$_2$(MoO$_4$)$_3$ leads to the orthorhombic Li$_2$Fe$_2$(MoO$_4$)$_3$ (mp-21778) which has also been experimentally reported alongside the Fe$_2$(MoO$_4$)$_3$ phase \cite{manthiram1987lithium}. In the deintercalated phase, Mo and Fe are in oxidation states +6 and +3, respectively, with Fe undergoing reduction to the +2 oxidation state upon Li intercalation. The constructed seawater Pourbaix diagram of this system is shown in \textbf{Figure~S7a}.

\subsubsection{Mg-selective}
\begin{figure}[h!]
\centering
\includegraphics[width=\textwidth]{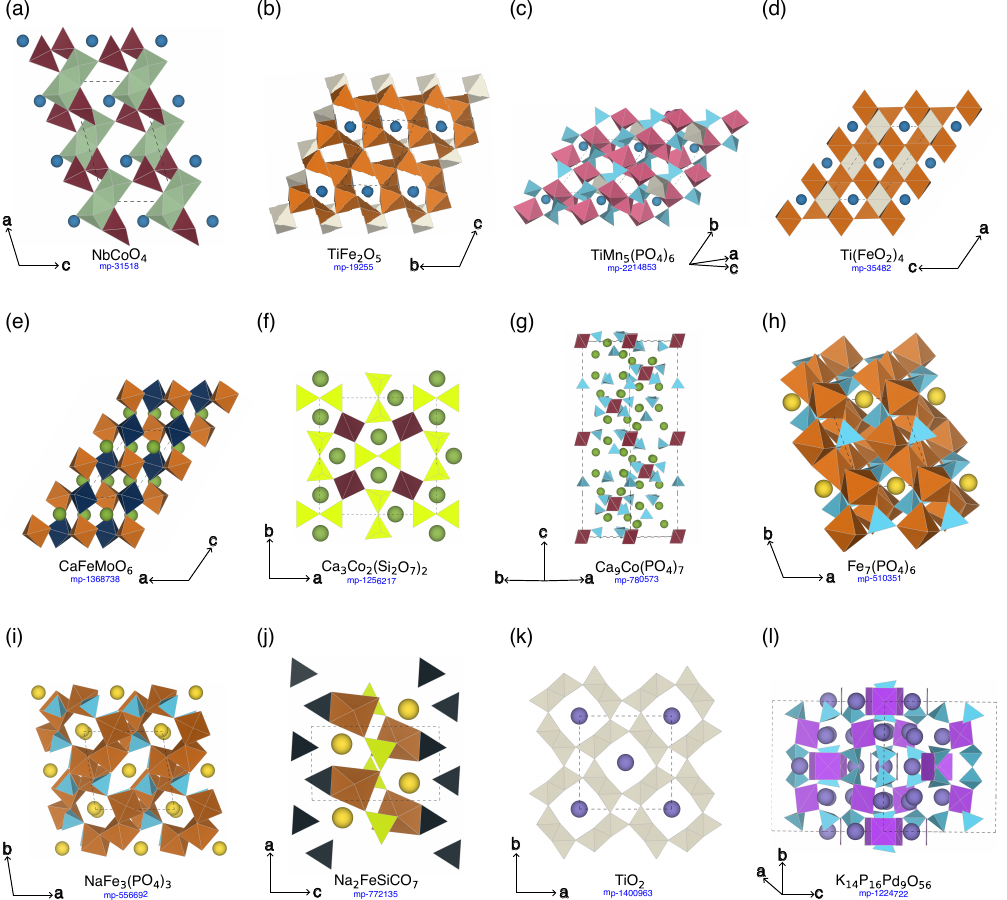}
\caption{(a)-(d) Mg-selective electrodes, where the blue spheres indicate possible Mg$^{2+}$ sites. (e)-(g) Ca-selective electrodes, where the green spheres indicate Ca${^{2+}}$. (h)-(j) Na-selective electrodes, with yellow spheres representing Na$^+$. (k)-(l) K-selective electrodes, with purple spheres representing K$^+$. The chemical formulae and MP ID below each panel corresponds to the most de-intercalated state represented in the corresponding graph (see \textbf{Figure~S11}). The dashed lines in each panel show the conventional unit cell. Following is the color scheme of the polyhedra: orange $-$ FeO$_6$ octahedra/FeO$_5$ trigonal bipyramids in panels (b), (d), (e), (h)-(j);  light-blue $-$ PO$_4$ tetraheadra in (c), (g)-(i), (l); light-gray $-$ TiO$_6$ octahedra in (b)-(d), (k); dark-brown $-$ CoO$_6$ octahedra in  (a), (f), (g); neon-green $-$ SiO$_4$ tetrahedra in (f), (j); olive-green $-$ NbO$_6$ octahedra in (a); pink $-$ MnO$_6$ octahedra in (c); dark-blue $-$ MoO$_6$ octahedra in (e); dark-gray $-$ trigonal planar CO$_3$ in (j); and bright-purple $-$ PdO$_6$ octahedra/PdO$_5$ square pyramids in (l).}
\label{fig:7}
\end{figure}

\noindent The most deintercalated versions of the Mg-selective hosts are summarized in panels a-d of \textbf{Figure~{\ref{fig:7}}} and their corresponding seawater Pourbaix diagrams are compiled in \textbf{Figure~S7d-g}. The experimentally observed hosts that are Mg-selective include NbCoO$_4$ (mp-31518, \textbf{Figure~{\ref{fig:7}a}}) and TiFe$_2$O$_5$ (mp-19255, \textbf{Figure~{\ref{fig:7}b}}). NbCoO$_4$ is a monoclinic structure ($C2/m$ spacegroup \cite{lehmann1980einkristalle}), which forms the theoretical Mg$_{1/2}$NbCoO$_4$ (mp-2219413) upon Mg intercalation. The deintercalated NbCoO$_4$ phase contains Nb and Co in the oxidation states +5 and +3, respectively, forming a network of edge sharing and corner sharing NbO$_6$ and CoO$_6$ octahedra. Mg$^{2+}$ intercalation results in the reduction of Co only to its oxidation  +2, which is accompanied by distortions in the CoO$_6$ octahedra. Mg occupies two symmetrically distinct triangular bipyramidal voids in the host. Similar to NbCoO$_4$, TiFe$_2$O$_5$ is also monoclinic ($C2/c$ spacegroup \cite{drofenik1981new}) and forms the theoretical Mg$_{1/2}$TiFe$_2$O$_5$ structure (mp-2229085, $P1$ spacegroup) upon Mg intercalation. The deintercalated structure consists of edge- and corner-sharing FeO$_6$ octahedra that are intertwined with corner-sharing TiO$_6$ octahedra, with Ti and Fe in their +4 and +3 oxidation states, respectively. With Mg intercalation, Fe reduces to its +2 oxidation state alongside distortions both in the FeO$_6$ and TiO$_6$ octahedra. 

There are two theoretical Mg-selective candidates that we identify in our workflow, namely, TiMn$_5$(PO$_4$)$_6$ (mp-2214853, \textbf{Figure~~{\ref{fig:7}c}}) and Ti(FeO$_2$)$_4$ (mp-35482, \textbf{Figure~~{\ref{fig:7}d}}). MgTiMn$_5$(PO$_4$)$_6$ (mp-2216952), which forms upon Mg intercalation in TiMn$_5$(PO$_4$)$_6$ shares the same triclinic $P1$ spacegroup with the deintercalated host, which contains TiO$_6$ octahedra, MnO$_6$ octahedra and PO$_4$ tetrahedra. Mg intercalation into the structure is expected to reduce Mn$^{3+}$ to Mn$^{2+}$ with Ti remaining in its +4 oxidation state. However, we do not expect the TiMn$_5$(PO$_4$)$_6$ to be a feasible candidate since the structure is relatively `closed' making it unlikely to facilitate diffusion of Mg$^{2+}$. In the case of Ti(FeO$_2$)$_4$, the structure is an ordered, monoclinic ($C2/m$ spacegroup) representation of an experimentally reported defective spinel structure that contains both Ti and Fe across tetrahedral and octahedral sites of the lattice \cite{xu1997transformation}. Specifically, Ti(FeO$_2$)$_4$ has Ti$^{4+}$ occupying only the octahedral sites of the spinel while Fe$^{3+}$ occupies both tetrahedral and octahedral sites. Mg intercalation yields MgTi(FeO$_2$)$_4$ (mp-2228463, $C2/m$), resulting in the reduction of Fe$^{3+}$ and the occupation of Mg within the octahedral voids. Given that spinels generally provide reasonable diffusivity for Mg$^{2+}$ \cite{canepa2017high,sai2017influence}, we expect Ti(FeO$_2$)$_4$ to be a potential candidate for experimental trials. 

\subsubsection{Ca-selective}
\noindent There are three candidate hosts that are Ca-selective per our workflow, which are compiled in panels e-g in \textbf{Figure~{\ref{fig:7}}} with their corresponding seawater Pourbaix diagrams listed in \textbf{Figure~S7h-j}. Among these Ca$_2$FeMoO$_6$ (mp-18783) and Ca$_2$CoSi$_2$O$_7$ (mp-18786) have been experimentally reported. Ca$_2$FeMoO$_6$ is a monoclinic double perovskite with $P2_1/n$ symmetry containing alternating FeO$_6$ and MoO$_6$ octahedra which are tilted to optimize coordination around the Ca$^{2+}$ that occupies the 'A' site of a ABO$_3$ perovskite \cite{alonso2000preparation}. The most deintercalaed version of Ca$_2$FeMoO$_6$ found in MP is the CaFeMoO$_6$ (mp-1323783, \textbf{Figure~~{\ref{fig:7}e}}), with another deintercalated structure, Ca$_{3/2}$FeMoO$_6$ (mp-1368738) are also found in MP. However, we find only the Ca$_{3/2}$FeMoO$_6$-Ca$_2$FeMoO$_6$ pair to be seawater metastable. 

The other experimental phase, Ca$_2$CoSi$_2$O$_7$ is a tetragonal structure ($P\bar{4}2_1m$ symmetry) with corner-sharing CoO$_4$ tetrahedra and Si$_2$O$_7$ pyrosilicate dimers \cite{hagiya1993modulated, ardit2012local, kusaka2001determination, torres2021enlisting, kimata1982four}. The most deintercalated version of this host in MP is Ca$_{3/2}$CoSi$_2$O$_7$ (listed as Ca$_3$Co$_2$(Si$_2$O$_7$)$_2$, mp-1256217, \textbf{Figure~{\ref{fig:7}f}}). Here Si remains in the +4 oxidation state and Co acts as the redox-active element with +2 and +3 oxidation states in the intercalated and deintercalated states, respectively. Ca$_9$Co(PO$_4$)$_7$ (mp-780573, \textbf{Figure~~{\ref{fig:7}g}}) is a theoretical deintercalated phase with a theoretical intercalated Ca$_{9.5}$Co(PO$_4$)$_7$ (mp-1228767) phase, where both structures exhibit trigonal ($R3c$) symmetry. The deintercalated phase contains Co in the +3 oxidation state forming CoO$_6$ octahedra, which reduce to Co$^{2+}$ upon Ca intercalation.

\subsubsection{Na-selective}
\noindent The most deintercalated versions of the Na-selective hosts are summarized in panels h-j of \textbf{Figure~~{\ref{fig:7}}}, with their seawater Pourbaix diagrams given in \textbf{Figure~S7k-m}. Note that Na-selective hosts may still be useful for desalination applications, since Na$^{+}$ is the most abundant and critical cation to be removed from seawater. Among the candidate hosts, Fe$_7$(PO$_4$)$_6$ (mp-510351, \textbf{Figure~~{\ref{fig:7}h}}) is an experimentally reported \cite{duan2004hydrothermal, vencato1994crystal, gorbunov1980crystal, kozlyakova2018tuning} triclinic structure with a $P\bar1$ spacegroup. The framework contains infinite chains of edge-sharing FeO$_6$ octahedra and edge-sharing FeO$_5$ trigonal bipyramids, which are linked by a set of corner-sharing FeO$_6$ octahedra and PO$_4$ tetrahedra. In the completely deintercalated structure, Fe exists in both its +2 and +3 oxidation states in a 3:4 ratio across all Fe-containing polyhedra \cite{belokoneva2003synthesis, sobolev2018magnetic, donni2026magnetic}. Addition of Na leads to an intercalated structure where the Fe$^{2+}$:Fe$^{3+}$ ratio increases to approximately  4:3, resulting in the formation of the experimentally reported NaFe$_7$(PO$_4$)$_6$ (mp-1205097) $P\bar1$ structure \cite{belokoneva2003synthesis, kozlyakova2018tuning}. Further optimization of the Na intercalation process can potentially enable more Na atoms to be captured by the structure, given that all Fe atoms are not fully reduced in NaFe$_7$(PO$_4$)$_6$.

NaFe$_3$(PO$_4$)$_3$ (mp-556692, \textbf{Figure~{\ref{fig:7}i}}) and its intercalated phase Na$_2$Fe$_3$(PO$_4$)$_3$ (mp-558822) both are experimentally reported alluaudite type structures \cite{yakubovich1977crystal, corbin1986synthesis} with monoclinic $C2/c$ spacegroup. The framework contains zigzag chains of edge-sharing FeO$_6$ octahedra with PO$_4$ tetrahedra forming bridges between the Fe-O chains via corner-sharing, leading to two distinct one dimensional Na diffusion pathways, typical of alluaudites \cite{deb2022critical}. While NaFe$_3$(PO$_4$)$_3$ has been reported only as a minor hydrothermal by-product \cite{corbin1986synthesis}, Na$_{2-x}$Fe$_3$(PO$_4$)$_3$ ($x \approx 0$ to $0.5$) is readily synthesizable via various routes, and has been used as a Na-ion cathode with non-aqueous electrolytes \cite{huang2015self, liu2017synthesis, essehli2020selective, walczak2019alluaudite}. We estimate that sodiation of NaFe$_3$(PO$_4$)$_3$ to Na$_2$Fe$_3$(PO$_4$)$_3$, leading to reduction of Fe$^{3+}$ to  Fe$^{2+}$, occurs at 1.04~V vs.~SHE (3.75~V vs.~Na/Na$^{+}$), which is $\sim$0.7-1.0~V above the experimentally reported intercalation voltages \cite{essehli2020selective, liu2017synthesis}. We attribute this discrepancy in the intercalation voltage to the different Na site occupancy of the NaFe$_3$(PO$_4$)$_3$ structure in MP \cite{corbin1986synthesis}, versus experimental Na$_{2-x}$Fe$_3$(PO$_4$)$_3$ structures \cite{liu2017synthesis, essehli2020selective}. 

Na$_3$Fe(CO$_3$)(SiO$_4$) (documented as Na$_3$FeSiCO$_7$, mp-771204) is a monoclinic mixed-polyanion silicate carbonate with $P2_1/m$ symmetry, similar to the sidorenkite Na$_3$Mn(PO$_4$)(CO$_3$) structure \cite{kurova1980crystal,
krivovichev2013refinement, chen2013sidorenkite}. It contains FeO$_6$ octahedra,  sharing four corners with SiO$_4$ tetrahedra and one edge with trigonal-planar CO$_3$. The most deintercalated version of Na$_3$Fe(CO$_3$)(SiO$_4$) found in MP is Na$_2$Fe(CO$_3$)(SiO$_4$) (mp-772135), as shown in \textbf{Figure~{\ref{fig:7}j}}. Unlike a phosphate carbonate (e.g., Na$_x$Fe(PO$_4$)(CO$_3$)), where the redox-active Fe is either in the +2 (intercalated) or +3 (deintercalated) oxidation states, replacing Si$^{4+}$  with P$^{5+}$ increases the Fe oxidation state to +3 in the sodiated state, and +4 in the desodiated state \cite{hautier2011novel}. Both the sodiated and desodiated silicate carbonates are theoretical structures, and Hautier et. al.'s \cite{hautier2011novel} high-throughput computational study reported Na$_3$Fe(CO$_3$)(SiO$_4$) (where Fe is in the +3 state) to be stable at 0~K. However, due to the presence of Fe in its +4 oxidation state with sodium removal, synthesis of this phase may be challenging.

\subsubsection{K-selective}
\noindent Both the shortlisted K-selective electrodes are theoretical, namely, TiO$_2$ (mp-1400963, \textbf{Figure~~{\ref{fig:7}k}}) and K$_{14}$P$_{16}$Pd$_{9}$O$_{56}$ (mp-1224722, \textbf{Figure~~{\ref{fig:7}l}}), with the corresponding seawater Pourbaix diagrams shown in \textbf{Figure~S7n-o2}.  The P$_{16}$Pd$_{9}$O$_{56}$ host contains all the Pd in a planar geometry, bonded to four O$^{2-}$, with PO$_4$ tetrahedra connecting three different PdO$_4$ units, resulting an interconnnected network of prismatic sites that are occupied by K$^{+}$. Upon K intercalation, there are two possible compositions with the host per MP, namely, K$_{14}$P$_{16}$Pd$_{9}$O$_{56}$ (mp-1224722) and K$_{15}$P$_{16}$Pd$_{9}$O$_{56}$ (mp-1212473). We expect the P$_{16}$Pd$_{9}$O$_{56}$ host to be challenging to realise experimentally, owing to the high cost of Pd. 

The TiO$_2$ host is tetragonal ($I4/m$ spacegroup) with a hollandite MnO$_2$-like structure \cite{zeng2025role}, and TiO$_6$ forming a mixture of corner- and edge-sharing octahedra that allow for one dimensional channels that accommodate K$^{+}$ within prismatic voids. Upon K intercalation, the TiO$_2$ host can form two intercalated structures, namely, K$_{1/16}$TiO$_2$ (mp-757418) and K$_{1/6}$TiO$_2$ (mp-765472), per MP. Although MP flags all three structures as theoretical, they correspond closely to the experimentally reported TiO$_2$, K$_{0.06}$TiO$_2$ and K$_{0.17}$TiO$_2$ \cite{latroche1989new, jo2023hollandite}. While Li intercalates into several TiO$_2$ polymorphs (\cite{liu2016recent, dylla2013lithium}), the experimentally reported Li$^{+}$ insertion voltage into the hollandite TiO$_2$ lies below the seawater stability window \cite{sakao2013lithium}. This unfavorable experimental Li-intercalation voltage, combined with the fact that MP does not contain any Li-intercalated hollandite-TiO$_2$ structures, indicates that hollandite-TiO$_2$ is likely to be selective towards K$^{+}$ in seawater conditions.

\subsection{Shortlisted non-selective hosts}
\label{subsec: Shortlisted non-selective hosts}
\noindent In this section, we provide detailed descriptions of the host structures that are identified as most promising non-selective hosts (i.e., capable of intercalating multiple cations), as compiled in \textbf{Figures~{\ref{fig:6}}a} and \textbf{S12}. The constructed seawater Pourbaix diagrams for all candidates described here are compiled in \textbf{Figures~S8-S10} of the SI, while the corresponding computational graphs are shown in \textbf{Figures~S13} and \textbf{{\ref{fig:3}}d}.\\

\subsubsection{Olivine-FePO$_4$}
While our workflow does not identify the ideal desalination electrode that can intercalate all five electroactive ions considered, the closest to the ideal candidate that we identify is olivine-FePO$_4$ (mp-20361, \textbf{Figure~{\ref{fig:6}}f}). While we do identify layered-CoO$_2$ (mp-1275937) and layered-FeO$_2$ (mp-25236) to also intercalate four out of five cations considered in this work, both hosts do not exhibit metastability regions that occupy near 80\% of the seawater stability window and hence are not feasible candidates. In terms of structure, olivine-FePO$_4$ contains parallel planes of equivalent corner-sharing FeO$_6$ octahedra, with Fe in its +3 oxidation state, that are bridged by PO$_4$ tetrahedra. The resulting framework provides one dimensional diffusion channels for the intercalating ions, with the reduction of Fe from +3 to +2 upon intercalation. We expect the FePO$_4$ to intercalate Li, Mg, Na, and K under seawater conditions, with the associated Pourbaix diagrams compiled in \textbf{Figure~S10} and the computational graph in \textbf{Figure~{\ref{fig:3}}d}.

Li and Na intercalation in olivine-FePO$_4$ are well documented with their reported intercalation voltages closely matching our predicted values of 0.46 and 0.29~V vs.~SHE, respectively \cite{wang2011olivine, zhu2013comparison, dong2011crystal, yan2022defining, fang2015high, ali2016polythiophene}. While Mg and K intercalated structures in MP are theoretical, there have been experimental reports of Mg \cite{shan2017olivine} and K \cite{ozdogru2022impact} intercalation in FePO$_4$ in aqueous media, with K intercalation reported to be irreversible due to amorphization caused by large lattice strains. We do show that intercalation of K in FePO$_4$, although possible, is not as (meta)stable as the intercalation of Li, Na or Mg, as indicated by the lower intersecting metastability area ($\sim$74\% of the seawater stability window) signifying the susceptibility of the system to decompose under high overpotentials. Also, experiments highlight limited capacity for intercalation of Mg in FePO$_4$, which agrees with our observation as well, given that the most intercalated metastable Mg phase is Mg$_{1/4}$FePO$_4$ that has Mg occupying only one-quarter of all possible intercalation sites. Interestingly, while there have been experimental reports of possible Ca intercalation in olivine-FePO$_4$ \cite{kim2020high}, there are no Ca containing olivine structures in MP, resulting in a lack of rigorous evaluation of Ca intercalation in olivine-FePO$_4$.

\subsubsection{Multi-cation intercalating hosts}
\noindent Hosts that can intercalate multiple cations simultaneously, such as two or three from the five cations considered in this work, can have use cases for different applications. For instance, Na$^{+}$ remains the critical entity to remove in any desalination process, so hosts that can remove Na$^{+}$ efficiently with any other cation can be useful for desalination workflows. Similarly, electrodes that can simultaneously remove Mg$^{2+}$ and Ca$^{2+}$ can be used for water softening applications. Moreover, potentiostatic techniques can be used to make non-selective hosts selective to a specific cation, if the intercalation voltages of the different cations are sufficiently far apart ($>$0.1~V, \textbf{Figure~{\ref{fig:6}}b}). Thus, similar intercalation voltages facilitate hosts to be non-selective while distinct intercalation voltages enable hosts to be selective. Moreover, intercalating multiple ions when their potentials are significantly separated may by challenging, since the reaction kinetics decay exponentially with the overpotential. Note that the similarities or differences in intercalation voltages with distinct cations in a structure are determined by two factors: $i$) relative (in)stability of the intercalated and deintercalated compositions (which can be modified with doping for instance), and $ii$) differences in the standard reduction potentials between the electroactive ions (cannot be modified).

In terms of hosts that can simultaneously intercalate two cations, at similar voltages, we identify maricite-FePO$_4$ (mp-776288, \textbf{Figure~S12g} \cite{ong2011voltage, kim2015unexpected}) to be useful in the simultaneous removal of Na (at 0.36~V vs.~SHE) and Li (at 0.28~V vs.~SHE), where only the Na-intercalated state is experimentally reported. Similarly, the high-pressure $\beta^{'}$-FePO$_4$ phase (mp-25303, \textbf{Figure~S12e} \cite{garcia2001influence}) can intercalate Li (at 0.36~V vs.~SHE) and Na (at 0.30~V vs.~SHE), with the Li-intercalated phase being experimentally reported. The monoclinic-FePO$_4$ polymorph (mp-752872, $P2_1/c$, \textbf{Figure~S12f}) that has FeO$_5$ bipyramidal units instead of the FeO$_6$ octahedra seen in the olivine, can theoretically intercalate both Ca and Mg at 0.28 and 0.30~V vs.~SHE, respectively. Fe$_2$(MoO$_4$)$_3$ (mp-1042928, \textbf{Figure~S12c}) is the final two ion intercalating host that we find with similar intercalation voltages of -0.057 and -0.141~V vs.~SHE for Mg and Ca, respectively, with all structures being theoretical in this system.

We do not find any seawater metastable electrodes that intercalate more than two different ions at a similar voltage. However, three hosts that show the ability to intercalate three ions, namely, Fe(PO$_3$)$_3$ (mp-25310, \textbf{Figure~S12i}), NaCoPCO$_7$ (mp-1101670, \textbf{Figure~S12l}), and LiFePCO$_7$ (mp-1176697, \textbf{Figure~S12k}), do exhibit similar voltages for the intercalation of two out of the three ions. For example, Fe(PO$_3$)$_3$ has an experimentally reported Li-intercalated structure, LiFe(PO$_3$)$_3$ (mp-504179 \cite{genkina1983crystal}) with an intercalation voltage of 0.64~V vs.~SHE, which is close to its theoretical Na intercalation voltage of 0.56~V vs.~SHE, but significantly different from the theoretical Mg intercalation voltage of 0.01~V vs.~SHE. Similarly, the CoPCO$_7$ host intercalates Li and Na at $\sim$1.1~V vs.~SHE, while K intercalation occurs at a different voltage of 0.81~V vs.~SHE. Na$_3$CoPCO$_7$ (mp-770948 \cite{chen2012synthesis}) is the only experimentally reported electrode in the CoPCO$_7$ host. Among all the intercalation reactions into CoPCO$_7$, only the sodiation of Na$_2$CoPCO$_7$ (mp-770988) yields coexisting phases that cover more than 80\% of the seawater metastability window. The analogous Li (de)intercalation falls just short at 79\% (\textbf{Figure~S9d1$\dagger$}), whereas the K reaction covers only 54\% (\textbf{Figure~S9d3$\ddagger$}), as indicated by the hatched and faded points, respectively, in \textbf{Figure~{\ref{fig:6}b}}.

The FePCO$_7$ host can intercalate Li, Na, and K, where LiFePCO$_7$ (mp-1176697), NaFePCO$_7$ (mp-769633) and KFePCO$_7$ (mp-769579) constitute the most deintercalated phases. Sodiation of the theoretical Na$_2$FePCO$_7$ (mp-1176399) phase forms the experimentally reported topotactic structure Na$_3$FePCO$_7$ (mp-22135, \cite{chen2012synthesis}) at $-$0.14~V vs.~SHE, which is close to the intercalation voltage of -0.2~V vs.~SHE of the theoretical phase  K$_3$FePCO$_7$ (mp-1575580). Other possible intercalation voltages with the FePCO$_7$ host, where the co-existing phase occupy 70 to 80\% of the seawater metastability area are at 0.73~V vs.~SHE, when Na$_3$FePCO$_7$ forms directly from the sodiation of NaFePCO$_7$, and at 1.04~V vs SHE when theoretical K$_2$FePCO$_7$ (mp-773453) forms upon K-intercalation of KFePCO$_7$. Interestingly, the presence of topotactic LiFePCO$_7$ (mp-1176697) and Li$_3$FePCO$_7$ (mp-25445) electrodes classifies FePCO$_7$ as a three-ion intercalating electrode, however both the Li-containing structures are not metastable under seawater operating conditions.

\section{Discussion}
\noindent Here, we used a high-throughput screening workflow (\textbf{Figure~{\ref{fig:1}}}) based on DFT-calculations, the conceptualization and construction of seawater Pourbaix diagrams, and the usage of modified metastability criteria and calculated intercalation voltages to identify electrode materials for electrochemical ion recovery (ion-selective hosts) and desalination (non-selective hosts) applications. Specifically, we screened through a set of 20,414 structures (\textbf{Figure~{\ref{fig:2}}}) containing Li, Mg, Ca, Na, or K from the MP project, evaluated their metastability and their ability to (de)intercalate under seawater conditions, and compared the structures containing different electroactive ions via computational graphs to identify 11 hosts that are ion-selective, 7 hosts that can intercalate two ions simultaneously, and the olivine-FePO$_4$ that can intercalate Li, Na, Mg, and K (\textbf{Figure~{\ref{fig:6}}}). Importantly, eight ion-selective and six non-selective (including olivine-LiFePO$_4$) electrode candidates that we identified as most promising exhibit at least one experimentally known composition/structure, highlighting their potential in experimental trials. A summary of the most promising candidates is compiled in \textbf{Table~{\ref{table:1}}}.

One of the approximations involved in our work is in the way we construct our seawater Pourbaix diagrams. Typically, Pourbaix diagrams obtain the thermodynamically stable phase(s) by minimizing the Pourbaix potential, which is open to H and O (i.e., the system is in equilibrium with an infinite reservoir of H and O atoms) at different pH and applied potential. Thus, the classical Pourbaix potential is strictly meant for modeling the stability of a given solid (such as a potential electrode material considered here) in pure water or dilute solutions (e.g., \textbf{Figure~{\ref{fig:2}}c}) and not a seawater medium that contains `large' reservoirs of non-H,O species, such as Cl, Na, Mg, S, etc. Indeed, including large reservoirs of non-H,O species within the Pourbaix potential requires the introduction of a chemical potential ($\mu$) for each species, which is non-trivial. For instance, $\mu_{\text{H}}$ and $\mu_{\text{O}}$ are tied to a constant $\mu_{\text{H}_2{\text{O}}}$ across pH and potentials  \cite{persson2012prediction, patel2019efficient}, allowing $\mu_{\text{H}}$ to be directly linked to pH (see \textbf{Section~S2}). But the same approximation cannot be made for $\mu_{\text{Cl}}$, since Cl can form Cl$^-$, ClO$_4^-$, HClO$_{4 \text{(s)}}$, or ClO$_{2 \text{(s)}}$ depending on pH and potential, meaning $\mu_{\text{Cl}}$ (or $\mu_{\text{HCl}}$) cannot be assumed to be a constant across pH/potential like $\mu_{\text{H}_2\text{O}}$. However, we are interested in finding intercalation electrodes that are stable in seawater and can capture cations, signifying that we cannot use the classical Pourbaix diagram \emph{as is} implemented in \texttt{pymatgen}.

To circumvent the challenge of describing $\mu$ for additional species in our Pourbaix potential, we introduced the following modifications: ($i$) included all non-H,O elements in the aqueous reservoir (such as Cl) as part of the `solid' also so that we can minimize the `standard' Pourbaix potential, ($ii$) restricted our analysis to the top four most abundant elements found in seawater to reduce computational complexity (\textbf{Section~S3}), ($iii$) calculated the decomposition energy of a phase by enumerating all combinations of reactants (where possible reactants vary as a function of pH and potential) that can react with a metastable electrode (\textbf{Section~S5}), and ($iv)$ characterize the reaction that yields the highest decomposition energy as the actual decomposition energy of a metastable electrode and evaluate against our 0.3~eV~atom$^{-1}$ threshold (\textbf{Sections~{\ref{sec:metastability}}}, \textbf{S6}, and \textbf{Figure~{\ref{fig:3}}}). Note that modification ($i$) listed above actually restricts our stability evaluation to a smaller chemical space since the ratios of all non-H,O elements are fixed in the combined solid+solution system to satisfy mass conservation. This limitation is partly compensated by modification ($iii$), where the enumeration of decomposition reactions allows for varying ratios of the species that are supposed to be already present in seawater to react with the metastable solid (electrode), and the tighter threshold used in modification ($iv$). However, we do acknowledge the imperfections of constructing seawater Pourbaix diagrams as done here and we will be developing a more rigorous thermodynamic formalism as part of follow-up work.

\begin{table*}
\small
  \caption{\ Summary of the most promising ion-selective and non-selective intercalation electrodes identified in this work.}
  \label{table:1}
  \newcommand{\rschead}[2][l]{\begin{tabular}[b]{@{}#1@{}}#2\end{tabular}}
  \begin{tabular*}{\textwidth}{@{\extracolsep{\fill}}llllllrr}
    \hline
    Host & Ion & \rschead{Deintercalated\\formula} & \rschead{Deintercalated\\MP ID} & \rschead{Intercalated\\formula} & \rschead{Intercalated\\MP ID} & \rschead{Voltage\\(V vs.\ SHE)} & \rschead[r]{Intersecting\\metastability\\area (\%)} \\
    \hline
    \\[2pt]
    \multicolumn{8}{c}{\textit{Ion-selective hosts}} \\[4pt]
    Fe$_{2}$(MoO$_{4}$)$_{3}$ & Li & Fe$_{2}$(MoO$_{4}$)$_{3}$$^{*}$ & mp-21925 & Li$_{2}$Fe$_{2}$(MoO$_{4}$)$_{3}$$^{*}$ & mp-21778 & 0.08 & 91.6 \\
    Fe$_{2}$(PO$_{4}$)$_{3}$ & Li & Li$_{3}$Fe$_{2}$(PO$_{4}$)$_{3}$$^{*}$ & mp-25993 & Li$_{5}$Fe$_{2}$(PO$_{4}$)$_{3}$ & mp-1177302 & $-$0.30 & 88.0 \\
    LiFeP$_{2}$O$_{7}$ & Li & LiFeP$_{2}$O$_{7}$$^{*}$ & mp-19294 & Li$_{2}$FeP$_{2}$O$_{7}$ & mp-779056 & 0.02 & 88.0 \\
    NbCoO$_{4}$ & Mg & NbCoO$_{4}$$^{*}$ & mp-31518 & MgNb$_{2}$(CoO$_{4}$)$_{2}$ & mp-2219413 & $-$0.12 & 87.3 \\
    Ti(FeO$_{2}$)$_{4}$ & Mg & Ti(FeO$_{2}$)$_{4}$ & mp-35482 & MgTi(FeO$_{2}$)$_{4}$ & mp-2228463 & 0.53 & 85.1 \\
    TiFe$_{2}$O$_{5}$ & Mg & TiFe$_{2}$O$_{5}$$^{*}$ & mp-19255 & MgTi$_{2}$(Fe$_{2}$O$_{5}$)$_{2}$ & mp-2229085 & $-$0.63 & 99.5 \\
    Ca$_{3}$Co$_{2}$(Si$_{2}$O$_{7}$)$_{2}$ & Ca & Ca$_{3}$Co$_{2}$(Si$_{2}$O$_{7}$)$_{2}$ & mp-1256217 & Ca$_{2}$CoSi$_{2}$O$_{7}$$^{*}$ & mp-18786 & 0.76 & 87.3 \\
    CaFeMoO$_{6}$ & Ca & Ca$_{3}$Fe$_{2}$(MoO$_{6}$)$_{2}$ & mp-1323783 & Ca$_{2}$FeMoO$_{6}$$^{*}$ & mp-18783 & $-$0.17 & 91.6 \\
    Na$_{2}$FeSiCO$_{7}$ & Na & Na$_{2}$FeSiCO$_{7}$ & mp-772135 & Na$_{3}$FeSiCO$_{7}$ & mp-771204 & 1.16 & 80.1 \\
    NaFe$_{3}$(PO$_{4}$)$_{3}$ & Na & NaFe$_{3}$(PO$_{4}$)$_{3}$$^{*}$ & mp-556692 & Na$_{2}$Fe$_{3}$(PO$_{4}$)$_{3}$$^{*}$ & mp-558822 & 1.04 & 88.0 \\
    TiO$_{2}$ & K & TiO$_{2}$ & mp-1400963 & KTi$_{16}$O$_{32}$ & mp-757418 & $-$0.25 & 99.5 \\
    & K & KTi$_{16}$O$_{32}$ & mp-757418 & KTi$_{6}$O$_{12}$ & mp-765472 & $-$0.53 & 99.5 \\
    
    \\[3pt]
    \multicolumn{8}{c}{\textit{Non-selective hosts}} \\[4pt]
    FePO$_{4}$ & Na & FePO$_{4}$ & mp-776288 & NaFePO$_{4}$$^{*}$ & mp-19226 & 0.36 & 88.0 \\
     & Li & FePO$_{4}$ & mp-776288 & LiFePO$_{4}$ & mp-831160 & 0.28 & 83.3 \\[3pt]
    FePO$_{4}$ & Li & FePO$_{4}$$^{*}$ & mp-25303 & LiFePO$_{4}$$^{*}$ & mp-18951 & 0.36 & 88.0 \\
     & Na & FePO$_{4}$$^{*}$ & mp-25303 & NaFePO$_{4}$ & mp-753672 & 0.30 & 88.0 \\[3pt]
    FePO$_{4}$ & Mg & FePO$_{4}$ & mp-752872 & MgFe$_{4}$(PO$_{4}$)$_{4}$ & mp-1234799 & 0.30 & 88.0 \\
     & Ca & FePO$_{4}$ & mp-752872 & CaFe$_{4}$(PO$_{4}$)$_{4}$ & mp-1233687 & 0.28 & 88.0 \\[3pt]
    Fe$_{2}$(MoO$_{4}$)$_{3}$ & Ca & Fe$_{2}$(MoO$_{4}$)$_{3}$ & mp-1042928 & Ca$_{3}$Fe$_{4}$(MoO$_{4}$)$_{6}$ & mp-1043101 & $-$0.06 & 81.2 \\
     & Mg & Fe$_{2}$(MoO$_{4}$)$_{3}$ & mp-1042928 & Mg$_{3}$Fe$_{4}$(MoO$_{4}$)$_{6}$ & mp-1042940 & $-$0.14 & 83.3 \\[3pt]
    Fe(PO$_{3}$)$_{3}$ & Li & Fe(PO$_{3}$)$_{3}$ & mp-25310 & LiFe(PO$_{3}$)$_{3}$$^{*}$ & mp-504179 & 0.65 & 80.8 \\
     & Na & Fe(PO$_{3}$)$_{3}$ & mp-25310 & NaFe(PO$_{3}$)$_{3}$ & mp-764111 & 0.56 & 87.3 \\
     & Mg & Fe(PO$_{3}$)$_{3}$ & mp-25310 & MgFe$_{4}$(PO$_{3}$)$_{12}$ & mp-2234300 & $-$0.01 & 84.4 \\[3pt]
    FePCO$_{7}$ & K & KFePCO$_{7}$ & mp-769579 & K$_{2}$FePCO$_{7}$ & mp-773453 & 1.04 & 79.4 \\
     & Na & NaFePCO$_{7}$ & mp-769633 & Na$_{3}$FePCO$_{7}$$^{*}$ & mp-22135 & 0.73 & 79.4 \\
     & Na & Na$_{2}$FePCO$_{7}$ & mp-1176399 & Na$_{3}$FePCO$_{7}$$^{*}$ & mp-22135 & $-$0.14 & 88.0 \\
     & K & K$_{2}$FePCO$_{7}$ & mp-773453 & K$_{3}$FePCO$_{7}$ & mp-1575580 & $-$0.20 & 87.3 \\[3pt]
    CoPCO$_{7}$ & Na & Na$_{2}$CoPCO$_{7}$ & mp-770988 & Na$_{3}$CoPCO$_{7}$$^{*}$ & mp-770948 & 1.12 & 83.0 \\
     & Li & Li$_{2}$CoPCO$_{7}$ & mp-25454 & Li$_{3}$CoPCO$_{7}$ & mp-25458 & 1.06 & 79.0 \\
     & K & KCoPCO$_{7}$ & mp-773502 & K$_{3}$CoPCO$_{7}$ & mp-1580470 & 0.81 & 54.2 \\[3pt]
    FePO$_{4}$ (olivine) & Li & FePO$_{4}$$^{*}$ & mp-20361 & LiFePO$_{4}$$^{*}$ & mp-19017 & 0.46 & 88.0 \\
     & Na & FePO$_{4}$$^{*}$ & mp-20361 & Na$_{2}$Fe$_{3}$(PO$_{4}$)$_{3}$$^{*}$ & mp-1195335 & 0.34 & 88.0 \\
     & Na & Na$_{2}$Fe$_{3}$(PO$_{4}$)$_{3}$$^{*}$ & mp-1195335 & NaFePO$_{4}$$^{*}$ & mp-746030 & 0.21 & 88.0 \\
     & Mg & FePO$_{4}$$^{*}$ & mp-20361 & MgFe$_{4}$(PO$_{4}$)$_{4}$ & mp-2234827 & $-$0.17 & 88.0 \\
     & K & FePO$_{4}$$^{*}$ & mp-20361 & KFePO$_{4}$ & mp-746071 & $-$0.25 & 70.4 \\
    \hline
  \end{tabular*}
  \par\vspace{2pt}
  {\footnotesize\raggedright
  Phases with an asterisk (*) denote experimental structures. Ti(FeO$_2$)$_4$ for Mg intercalation, Na$_2$FeSiCO$_7$ for Na intercalation, and TiO$_2$ for K intercalation are the pure theoretical ion-selective hosts listed here. Similarly, Fe$_2$(MoO$_4$)$_3$ and FePO$_4$ (mp-725872), both for Mg-Ca intercalation, are the pure theoretical non-selective hosts. All voltages are in V vs.~SHE.\par}
\end{table*}

Besides the modifications to the Pourbaix potential, we assume an electrode to be ion-selective if MP does not contain any structure with another electroactive ion intercalated in the same host. As a consequence, the lack of data in MP could yield false positives, where an electrode seems to be ion-selective due to the absence of data rather than its intrinsic behavior. However, we expect the proposed experimentally known ion-selective electrodes (\textbf{Figure~{\ref{fig:6}}a} and \textbf{Table~{\ref{table:1}}}) to be more reliable than the theoretical electrodes (yellow sets of \textbf{Figure~{\ref{fig:4}}}), as experimentally known electrodes are more likely to have been explored with alternate electroactive ions, where the lack of reported structures with other ions possibly indicates the tendency of the host to be ion-selective. Note that the set of hosts that can theoretically intercalate multiple ions but do not exhibit the required metastability (i.e., systems with faded symbols in \textbf{Figure~{\ref{fig:6}}b}), such as Fe$_2$(SiO$_3$)$_2$ (mp-1253019, mp-25240) and Fe$_2$(PO$_4$)$_3$ (mp-26582), can be considered as potentially ion-selective candidates as well.

Among the initial set of shortlisted candidates in panels a and b of \textbf{Figure~{\ref{fig:6}}}, we observe the recurrence of O, Fe, and P across several hosts. While the presence of O can be attributed to its abundance in the aqueous reservoir, the frequent occurrence of Fe and P can provide crucial insights into the chemistry of seawater (meta)stable electrodes. For instance, \textbf{Figure~S15b} shows that the final set of shortlisted candidates (including theoretical hosts) has a significantly greater fraction of structures that contain Fe (74.2\%) and P (72\%), as compared to their fractions in the full dataset in (32.9\% Fe and 33.4\% P, \textbf{Figure~S15a}). A phenomenological explanation for the occurrence of P and Fe can be attempted based on their individual seawater Pourbaix diagrams, as shown in \textbf{Figure~S14a,b}. Dark-green (light-green) regions in \textbf{Figure~S14} contain at least one solid phase (only aqueous species). In the seawater Pourbaix diagram of Fe (\textbf{Figure~S14a}) we mostly observe dark-green regions that can be attributed to stable Fe$_2$O$_3$ (explaining the occurrence of Fe-containing candidates), while light-green regions appear only at lower potentials where Fe dissolves into FeOH$^{+}$. On the other hand, most of the seawater stability window is light-green for P, indicating presence of only aqueous species (\textbf{Figure~S14b}), which does not explain our observation of the occurrence of P among candidate seawater metastable electrodes. This indicates that the occurrence of P can not be explained by considering single element Pourbaix diagrams alone.

An examination of the combined Fe-P seawater Pourbaix diagram  (\textbf{Figure~S14c}) offers a better explanation on the frequent occurrence of Fe and P among candidate electrodes compared to the individual Fe and P Pourbaix diagrams. For example, the light-green regions of similar voltages ($-0.5$ to $-0.2$V) and pH ($7$ to $8.4$), where Fe and P individually dissolve to form FeOH$^{+}$ and HPO$_4^{2-}$, the combined Fe-P system forms a dark-green region with a stable Fe$_3$(PO$_4$)$_2$ solid. Thus, while P by itself dissolves, the presence of Fe leads P to form a stable solid phase. Hence, the combined presence of Fe and P widens the (meta)stability of any electrode phase that forms Fe$_3$(PO$_4$)$_2$ as a decomposition product, which is not possible with the sole presence of either Fe or P in the electrode, explaining the frequency of occurrence of both Fe and P in several candidate electrodes. Further, we observe P and Si to be the most commonly ocurring element in candidate electrodes when considering the presence of a transition metal (see \textbf{Figure~{\ref{fig:SI-14 compositional trends in metastable electrodes}}}), highlighting the solid-stabilizing effect of P in multi-component seawater systems.

While our workflow identifies olivine-FePO$_4$ as a promising desalination electrode (\textbf{Figure~{\ref{fig:6}}}), we do not identify the experimentally explored spinel-Mn$_2$O$_4$ as a candidate host for electrochemical ion capture, which can largely be attributed to the tendency of Mn to dissolve in aqueous media \cite{srimuk2020charge, ouyang2009jahn, jiang2022facet, liang2023enabling}. For example, we do find that Li intercalation in MnO$_2$ is feasible in seawater conditions (\textbf{Figure~S5b}), but the area of metastability of the MnO$_2$ host is only 25\% of the seawater stability window, thus making it susceptible to overpotentials. The tendency of MnO$_2$ to dissolve is primarily due to the formation of stable MnO$_4^{-}$ species at higher potentials and Mn$^{2+}$ at lower potentials, within the seawater stability window. Also, the presence of P with Mn (\textbf{Figure~S14d}), unlike the case with Fe, does not cause any stabilizing effect via the formation of a stable and solid Mn-based phosphate. Thus, Mn dissolution driving a reduction in the metastability area is the primary reason that Mn-containing hosts are not shortlisted in our screening.

Other approximations in our workflow include considering only aqueous species (i.e., ions and complexes) that contain a single non-H,O element, as constrained by the availability of experimental aqueous species data in MP \cite{persson2012prediction}, which excludes complexes such as NaSO$_4^-$. Additionally, all solid energies used in our work are DFT-calculated, using periodic boundary conditions, signifying that configurationally disordered, amorphous, and defective solids have not been considered. Further, errors associated with DFT calculations, such as deficiencies in the description of electronic exchange and correlation in solids, affect our predictions as well. Also, we approximate the activities of the aqueous species by their corresponding molar concentrations to obtain the Pourbaix potential of the dissolved species. Thus, a non-H,O element dissolving from a solid has to be equilibrated with respect to a constant pre-determined concentration in solution, which are the respective seawater concentrations for Cl, Na, Mg, S, Ca, and K and the dilute limit of 10$^{-6}$~M for all other elements. 

Also, we do not explore the thermodynamic and/or kinetic effects of co-intercalation of multiple ions, majorly due to the lack of available co-intercalated structures. Thus, we only report whether co-intercalation is possible within an electrode under seawater conditions (\textbf{Figure~{\ref{fig:6}}b}). We do observe a high number of Li-based candidates in our screening (\textbf{Figure~{\ref{fig:2}}b}), which can largely be attributed to the bias in our dataset that itself derives from the extensive research done on LIBs. Indeed, the number of queried phases are in the sequence Li$>$Mg$>$Na$>$Ca$>$K, while the shortlisted candidates follow the order Li$>$Mg$>$Ca$>$Na$>$K (\textbf{Figure~{\ref{fig:4}}}). In any case, we expect all our predicted non-Li intercalating structures to not intercalate any Li at all, given that it is more likely for prior studies to attempt Li intercalation than other ionic intercalation in any host. Furthermore, we do not evaluate kinetic properties of hosts, such as the activation barrier for ionic migration of the electroactive ion(s), which can play a role in the practical feasibility of using an electrode for ion capture.

\section{Conclusion}
\noindent In summary, we have used a high-throughput DFT-based screening approach to identify novel intercalation electrodes for electrochemical ion recovery and desalination, which are crucial technologies that can aid in our transition to a sustainable energy future and better use of our available water sources. Upon querying 20,414 structures containing Li, Mg, Ca, Na, or K along with redox-active TM(s) and anion(s) from the MP, we developed a four-step screening workflow involving $i$) the construction of seawater Pourbaix diagrams to include the effects of ions normally present in seawater, $ii$) the calculation of the decomposition energy of metastable electrodes systematically using the corresponding seawater Pourbaix diagrams, $iii$) the determination of the metastability of (de)intercalated structures based on a modified ($<$0.3~eV~atom$^{-1}$) threshold, $iv$) the evaluation of average intercalation voltages and intersecting metastability area for deintercalated-intercalated pairs of structures (to be $>$80\% of the seawater stability window), and $v$) the identification of promising ion-selective and non-selective candidates via the construction of computational graphs to encode selectivity across hosts. Importantly, we identified eight experimentally known hosts and three theoretical hosts to be promising ion-selective candidates, four experimentally known hosts that can intercalate two ions simultaneously, and olivine-FePO$_4$ as the closest candidate to an ideal desalination electrode. Some of the promising materials identified in our workflow, which have not been explored yet for selective ion recovery from aqueous media, include, Fe$_2$(MoO$_4$)$_3$, Li$_3$Fe$_2$(PO$_4$)$_3$, LiFeP$_2$O$_7$ for Li, NbCoO$_4$ and TiFe$_2$O$_5$ for Mg, Ca$_3$Co$_2$(Si$_2$O$_7$)$_2$ and Ca$_3$Fe$_2$(MoO$_6$)$_2$ for Ca, Fe$_7$(PO$_4$)$_6$ and NaFe$_3$(PO$_4$)$_3$ for Na, and hollandite-TiO$_2$ for K. We are hopeful that our high-throughput screening framework, qualitative trends and thermodynamic insights, and the final list of identified candidates accelerates the practical development of electrochemical ion capture technologies.

\section*{Acknowledgments}
\noindent G.S.G. acknowledges funding by the Science and Engineering Research Board, Government of India, under the sanction number IPA/2021/000007. S.P. thanks the Ministry of Human Resource Development, Government of India, for financial assistance. A.K., M.H., and H.G.S. acknowledge funding by the German Federal Ministry of Research, Technology and Space (BMFTR) and the Ministry of Economic Affairs, Industry, Climate Action and Energy of the State of North Rhine-Westphalia through the project HC-H2. M.H. and H.G.S. acknowledge funding from the German Federal Ministry of Research, Technology and Space (BMFTR) via project 05K24CJ1. A.K. thanks the Stiftung Stipendien-Fonds of the German Chemical Industry Association (Verband der Chemischen Industrie, VCI) for a Kekulé fellowship. S.P. and G.S.G. acknowledge the computational resources provided by the Supercomputer Education and Research Centre, IISc, for enabling some of the calculations showcased in this work. The authors gratefully acknowledge the computing time provided to them on the high-performance computers noctua1 and noctua2 at the NHR Center PC2. This was funded by the Federal Ministry of Education and Research and the state governments participating on the basis of the resolutions of the GWK for national high-performance computing at universities (www.nhrverein.de/unsere-partner). The computations for this research project were performed using computing resources under projects hpc-prf-emdft and hpc-prf-desal. The authors also gratefully acknowledge the J\"{u}lich Supercomputing Centre at Forschungszentrum J\"{u}lich, Germany, for access to the JURECA supercomputer under the project `desal', which enabled some of the calculations reported in this work.

\section*{Data and code availability}
\noindent All calculated data and the scripts used in this work are available freely online on our \href{https://github.com/sai-mat-group/SPICES}{GitHub} repository.

\section*{Conflicts of interest}
\noindent The authors have no conflicts of interest to declare.

\bibliographystyle{unsrt}
\bibliography{references}

\end{document}